\documentclass[twocolumn, twocolappendix]{aastex631}

\usepackage{amssymb}
\usepackage{amsmath}

\newcommand{\vect}[1]{\boldsymbol{#1}}
\newcommand{\dvect}[1]{\dot{\boldsymbol{#1}}}
\newcommand{\uvect}[1]{\vect{\hat{#1}}}

\shorttitle{Formation of WASP107 b }
\shortauthors{Yu and Dai}

\begin{document}

\title{Are WASP-107-like Systems Consistent with High-eccentricity Migration?}

\author[0000-0002-6011-6190]{Hang Yu}
\affiliation{eXtreme Gravity Institute, Department of Physics, Montana State University,
Bozeman, Montana 59717, USA}
\email{hang.yu2@montana.edu}

\author[0000-0002-8958-0683]{Fei Dai}
\affiliation{Institute for Astronomy, University of Hawai`i, 2680 Woodlawn Drive, Honolulu, HI 96822, USA}
\email{fdai@hawaii.edu}




\begin{abstract}
WASP-107 b seems to be a poster child of the long-suspected high-eccentricity migration scenario. It is on a 5.7-day, polar orbit. The planet is Jupiter-like in radius but Neptune-like in mass with exceptionally low density. WASP-107 c is on a 1100-day, $e=0.28$ orbit with at least Saturn mass. Planet b may still have a residual eccentricity of $0.06\pm 0.04$: the ongoing tidal dissipation leads to the observed internally heated atmosphere and hydrodynamic atmospheric erosion. We present a population synthesis study coupling octupole Lidov-Kozai oscillations with various short-range forces, while simultaneously accounting for the radius inflation and tidal disruption of the planet. We find that a high-eccentricity migration scenario can successfully explain nearly all observed system properties. 
Our simulations further suggest that the initial location of WASP-107 b at the onset of migration is likely within the snowline ($<0.5\,{\rm AU}$). 
More distant initial orbits usually lead to tidal disruption or orbit crossing. 
WASP-107 b most likely lost no more than 20\% of its mass during the high-eccentricity migration, i.e. it did not form as a Jupiter-mass object. More vigorous tidally-induced mass loss leads to disruption of the planet during migration. We predict that the current-day mutual inclination between the planets b and c is substantial: at least 25-55$^\circ$ which may be tested with future {\it Gaia} astrometric observations. Knowing the current-day mutual inclination may further constrain the initial orbit of planet b. We suggest that the proposed high-eccentricity migration scenario of WASP-107 may be applicable to HAT-P-11, GJ-3470, HAT-P-18, and GJ-436 which have similar orbital architectures. 
\end{abstract}



\section{Introduction} 
\label{sec:intro}

Thanks to its large scale height and relatively bright host star (V=11.5, J=9.4), WASP-107 b is arguably one of the best characterized exoplanets to date. The host star is a K6 dwarf \citep[$T_{\rm eff} = 4300K$,][]{Anderson} with a rotation period of 17 days. Coupled with a high chromospheric S-index of 0.90 \citep{Baliunas}, the star could be as young as 600-Myr-old \citep[see also ][who suggested a more mature age if magnetic spin-down is stalled]{Piaulet}. See also \citet{Bouma_gyro}. Planet b has a Jupiter-like radius of 0.948$\pm0.03 R_j$ \citep{Dai_2017} and a Neptune-like mass of $30.5\pm1.7M_\oplus$ \citep{Piaulet}, hence an exceptionally low density of $\sim$0.1 g~cm$^{-3}$. The planet is on a 5.7-day orbit with a non-zero eccentricity of 0.06$\pm0.04$ \citep{Piaulet}. There is also a distant companion on a roughly 1100-day orbit with a projected mass $M_c $sin$~i$=0.36$M_{j}$. WASP-107 b was found to be on a polar, slightly retrograde orbit with a sky-projected stellar obliquity $I_{iS}={118^{+38}_{-19}}^\circ$ \citep{Dai_2017,Rubenzahl}. WASP-107 b is also vigorously losing its atmosphere as revealed by metastable Helium observations \citep{Spake,Allart,Kirk}. The mass loss rate is estimated to be about $1M_\oplus/$Gyr \citep{Wang_Dai}, and a stronger-than-solar stellar wind collimates the outflow into a comet-like tail \citep{Spake2021}. More recently, \citet{Dyrek} reported transmission spectroscopic observations of WASP-107 b using the Mid-Infrared Instrument (MIRI) of the James Webb Space Telescope \citep[JWST;][]{Gardner}. They reported clear evidence of photochemistry driven by high-energy irradiation from the host star that produced SO$_2$ \citep[see also][]{Tsai}. There are also hints of high-altitude ($10^{-5}$ bar) silicate cloud that demands vigorous vertical mixing. \citet{Welbanks:24} performed a panoramic analysis of all available transmission spectroscopy of WASP-107 b from Hubble \citep{Kreidberg}, MIRI \citep{Dyrek}, and JWST/NIRCam. \citet{Welbanks:24} suggest that WASP-107 b must be internally heated with an intrinsic temperature $T_{\rm int}>345K$. They also found a strong vertical mixing rate 
$\log [K_{zz}/(\rm cm^2\,s^{-1})] = 8.4-9$ on WASP-107 b. \citet{Welbanks:24} further reported an atmospheric metallicity that is 10-18 $\times$ solar metallicity, and a carbon-to-oxygen ratio C/O = 0.33$^{+0.06}_{-0.05}$. \citet{Sing} found similar conclusions using JWST/NIRSpec data. They reported even higher planetary atmospheric metallicity of 43$\pm$8 times solar, higher $T_{\rm int}=460\pm40K$, and stronger vertical mixing $\log [K_{zz}/(\rm cm^2\,s^{-1})] = 11.6\pm0.1$. Comparing with the host star, \citet{Hejazi} reported stellar abundance [Fe/H] = -0.071$\pm$0.014 and stellar C/O = 0.50$\pm$0.10.

We propose in this paper that a high-eccentricity migration (high-e migration) scenario \citep[e.g.,][]{Fabrycky:07, Dawson:18} can thread many observations of the WASP-107 system into a coherent story. WASP-107 b likely initially formed further out in the disk where conditions are more conducive for the accretion of a thick envelope \citep{Lee_accrete}. Secular evolution, particularly Lidov-Kozai (LK) oscillation~\citep{Naoz:16}, induced by planet c launched planet b into an eccentric, mutually inclined, high-stellar-obliquity \citep[${I_{iS}=118^{+38}_{-19}}^\circ$;][]{Rubenzahl} orbit around the host star. Eccentricity tidal interaction with the host star dissipates energy from the planet's orbit, causing it to slowly circularize while the semi-major axis decreases. Today, the planet has not fully circularized yet \citep[$e=$0.06$\pm0.04$;][]{Piaulet}. Orbital energy is still being dissipated within the planet and leads to significant internal heating ($T_{\rm int}>345K$, \citealt{Welbanks:24}; $T_{\rm int}=460\pm40K$, \citealt{Sing}). The internally heated planet coupled with high-energy irradiation from a young host star drives a vigorous atmospheric erosion on the planet  \citep[$1M_\oplus/$Gyr]{Wang_Dai,Spake2021}. This hydrodynamic outflow naturally provides the necessary vertical mixing to elevate silicate clouds \citep[$10^{-5}$ bar][]{Dyrek} to high altitudes \citep[see also][]{Wang_Dai_superpuff}.

 WASP-107 is not alone, such a high-e migration scenario may be widely applicable to several hot Neptune planets with similar orbital architectures e.g. HAT-P-11 \citep{Bakos,Sanchis_hat,Allart_hat,Xuan}, GJ-3470 \citep{Bonfils,Stefansson_3470,Lampon}, GJ-436 \citep{Butler,Ehrenreich,Bourrier_436}, HAT-P-18 \citep{Hartman,Esposito,Paragas}, and Kepler-1656 \citep{Angelo:22}. These planets have polar orbits around their host star; have residual orbital eccentricities; possess low-density atmospheres that may be eroding; and have more distant stellar or planetary companions (if enough long-term radial velocity monitoring is available).

In this work, we simulate directly the suspected high-e formation pathway for WASP-107 b with the aim of answering the following questions:

\begin{itemize}
     \item Can the observed planet c induce a high-e migration of planet b? What must be the initial mutual inclination between the two planets?  How about the current mutual inclination which may be constrained with astrometric measurements \citep{Gaia}?
    
    \item Can a high-e migration successfully reproduce the current-day polar orbit of WASP-107 b around the host star? 
    
    \item With a 5.7-day orbit around the host star, is WASP-107 b expected to be fully circularized or still experiencing eccentricity tides?

    \item Did WASP-107 b start out beyond the snowline of the disk before the migration? Or was the progenitor closer in?

    \item Was WASP-107 b born with a low mass ($\sim 0.1 M_J$), or could it have lost a significant portion of its mass during high-e migration?

    \item When was the radius WASP-107 b inflated? Could it survive tidal disruption with an inflated radius during the migration? 

\end{itemize}

The paper is structured as follows. In Sec.~\ref{sec:methods}, we lay out the numerical setup of our simulations. We present a few case studies where high-e migration successfully produced analogs of the WASP-107 system in Sec.~\ref{sec:results_examples}. We show our population synthesis results in Sec.~\ref{sec:results_pop_level}.  Finally, we conclude the findings of this paper in Sec.~\ref{sec:conclusions}

\section{Simulation Setup}
\label{sec:methods}
Our simulation leverages the well-characterized orbital architecture and planetary properties of the WASP-107 system. We stick to the observed properties of the system as much as we can, which fixes the mass and radius of the host star, the semi-major axis of the outer orbit. The key parameters we will explore are listed in Table \ref{tab:pop_syn_par}. $M_\ast$, $M_p$, and $M_c$ respectively represent the masses of the host star, the inner planet (WASP-107 b), and the outer planet (WASP-107 c). The Keplerian orbital elements of the inner planet b are without a subscript  (e.g. $a$ and $e$)  and those for the outer planet c have a subscript ``$o$''. We also use subscripts ``init'' and ``fin'' to indicate parameters at the beginning and the end of our simulations.

\subsection{Lidov-Kozai oscillation including octupole terms}
\label{sec:LK}

We evolve numerically the $M_\ast$-$M_p$-$M_c$ triple system under the Lidov-Kozai (LK) oscillation \citep[also known as the von-Zeipel-Lidov-Kozai,][]{Kozai,Lidov,Zeipel}. We also include short-range forces due to general relativity, tides, and rotation-induced quadrupole of the star and the planet. Our numerical setup closely follows the pioneering work of \citet{Anderson:16}; see also \citet{Vick:19}. A notable improvement is that we incorporated the back-reaction on the outer orbit at the octupole order based on \citet{Liu:15}. We will compare and contrast the simulations with and without including the octupole LK terms \citep{Naoz:11, Naoz:12, Naoz:13, Naoz:16, Li:14a, Li:14b, Albrecht:12, Antognini:15, Petrovich:15, Petrovich:16, Storch:17, Stephan:16, Stephan:17, Stephan:18, Liu:15, Liu:18, Anderson:16, Vick:19}.
Simulations evolved with the octupole terms are dubbed ``octLK'', and those including only quadrupole terms ``quaLK''. In both cases, we average over both the inner and outer orbits. Also incorporated in our simulation are mass loss from the planet and radius inflation; see respectively Sec.~\ref{sec:mass_loss} and Sec.~\ref{sec:R_infl}. 

Before proceeding, we briefly review the properties of the LK dynamics, which can be largely derived analytically when restricting to the quadrupole order. Conservation of total angular momentum and total energy (to the quadrupole order) leads to a relation \citep{Liu:15, Anderson:16, Liu:18}
\begin{align}
    &\frac{3}{8} \frac{e_{\rm max}^2}{j_{\rm min}^2}\left[5 \left(c_{io} + \frac{\eta}{2}\right)^2 
    - \left(3 + 4\eta c_{io} + \frac{9}{4}\eta^2\right)j_{\rm min}^2 + \eta^2 j_{\rm min}^4 \right]\nonumber\\
    &+\epsilon_{\rm GR}\left(\frac{1}{j_{\rm min}} - 1\right) 
    + \frac{\epsilon_{\rm ST}}{15} \left(\frac{1 + 3 e_{\rm max}^2 + \frac{3}{8}e_{\rm max}^4}{j_{\rm max}^9} - 1\right)=0,
    \label{eq:c_io_vs_j_min}
\end{align}
where $c_{io} = \cos I_{io, \rm{init}}$ is the cosine of the initial inclination between the inner and outer orbit, and $e_{\rm max} = \sqrt{1-j_{\rm min}^2}$. The smallest pericenter separation can be calculated from $e_{\rm max}$ as $r_{p,{\rm min}} = a (1 - e_{\rm max})$. Following \citet{Liu:18}, we have introduced 
\begin{align}
    \eta &= \left(\frac{L}{L_o}\right)_{e=0} = \frac{\mu}{\mu_o}\left[\frac{(M_\ast + M_p)a}{(M_\ast + M_p + M_c) a_o \left(1 - e_o^2\right)}\right]^{1/2}, \nonumber \\
    &= 0.052 \left(\frac{\mu}{0.12\,M_\odot}\right) \left(\frac{\mu_o}{0.8\,M_\odot}\right)^{-1} \left(\frac{M_\ast + M_p}{M_\ast + M_p + M_c}\right)^{1/2} \nonumber \\
    &\times \left(\frac{a}{0.2\,{\rm AU}}\right)^{1/2} \left(\frac{a_o(1-e_o^2)}{1.7\,{\rm AU}}\right)^{1/2},
\end{align}
to characterize the strength of the back reaction of the inner orbit on the outer one, with $\mu = M_p M_\ast/(M_\ast + M_p)$ and $\mu_o = M_c (M_\ast + M_p)/(M_\ast + M_p + M_c)$.
Also introduced are
\begin{align}
    \epsilon_{\rm GR} &= \frac{3 G (M_\ast + M_p)^2 a_{\rm o, eff}^3}{c^2 M_c a^4 }, \nonumber \\
    &=6.4\times 10^{-2} \left(\frac{M_\ast + M_p}{0.69\,M_\odot}\right)^2 \left(\frac{M_c}{0.8\,M_j}\right)^{-1} \nonumber \\
    &\times \left(\frac{a_{\rm o, eff}}{1.77\,{\rm AU}}\right)^{3} \left(\frac{a}{0.2\,{\rm AU}}\right)^{-4},  
    \label{eq:epsilon_GR}
    \\
    \epsilon_{\rm ST} &= \frac{15 k_{2p} M_\ast (M_\ast + M_p) a_{\rm o, eff}^3 R_p^5}{M_p M_c a^8}, \nonumber \\
    &= 1.2\times10^{-3} \left(\frac{k_{2p}}{0.37} \right)\left(\frac{M_p}{0.12\,M_j}\right)^{-1} \left(\frac{R_p}{0.94\,R_j}\right)^5 \nonumber \\
    &\times \left(\frac{M_\ast}{0.69\,M_\odot}\right) \left(\frac{M_\ast + M_p}{0.69\,M_\odot}\right) \left(\frac{M_c}{0.8\,M_j}\right)^{-1} \nonumber \\
    &\times  \left(\frac{a}{0.2\,{\rm AU}}\right)^{-8} \left(\frac{a_{\rm o, eff}}{1.77\,{\rm AU}}\right)^3, 
    \label{eq:epsilon_ST}
\end{align}
to characterize the strength of short-range forces due to general relativity (GR) and static tide (ST). In the equations above, $a_{\rm o, eff}=a_o\sqrt{1-e_o^2}$ and $a$ is evaluated at its initial value $a_{\rm init}$ (as we ignore tidal dissipation for now and assume energy is conserved). Numerical values are provided for typical parameters that describe the WASP-107 system. 
In the limit where $\eta=\epsilon_{\rm GR}=\epsilon_{\rm ST}=0$, Eq.~(\ref{eq:c_io_vs_j_min}) reduces to the well-known condition $e_{\rm max} = \sqrt{1-(5/3)c_{io}^2}$.

It is also instructive to consider the limiting eccentricity (or $j_{\rm lim}^2=1-e_{\rm lim}^2$) reached when only one of $(\epsilon_{\rm GR}, \epsilon_{\rm ST})$ is non-zero and the back-reaction $\eta$ is ignored. Approximately, we can set $e_{\rm max}\simeq 1$ and $c_{io}\simeq 0$, leading to a simple relation \citep{Anderson:16}
\begin{equation}
    \frac{\epsilon_{\rm GR}}{j_{\rm lim}} + \frac{7 \epsilon_{\rm ST}}{24 j^9_{\rm lim}}\simeq \frac{9}{8}.
\end{equation}
When general relativity dominates, we have
\begin{align}
    j_{\rm lim}^2|_{\rm GR} &\simeq \frac{64}{81}\epsilon_{\rm GR}^2 \nonumber \\
    &\simeq 3.2 \times 10^{-3} \left(\frac{a}{0.2\,{\rm AU}}\right)^{-8},
    \label{eq:j_lim_GR}
\end{align}
while the static tide-dominated case leads to
\begin{align}
    j_{\rm lim}^2|_{\rm ST} &\simeq \left(\frac{7}{27} \epsilon_{\rm ST}\right)^{2/9} \nonumber \\
    &\simeq 0.17 \left(\frac{a}{0.2\,{\rm AU}}\right)^{-16/9}. \label{eq:j_lim_ST}
\end{align}
The same numerical values used in Eqs.~(\ref{eq:epsilon_GR}) and (\ref{eq:epsilon_ST}) are adopted to arrive at the numerical estimations of $j_{\rm lim}$, and we have only kept the scaling with respect to $a$ explicitly as it represents one of the largest uncertainty of the system and we will aim to constrain it in Sec.~\ref{sec:pop_a_init}. As the greater of Eqs.~(\ref{eq:j_lim_GR}) and (\ref{eq:j_lim_ST}) determines $j_{\rm lim}$, it is typically the static tide that limits the maximum eccentricity excitation for WASP-107-like systems to around $e_{\rm max}\simeq 0.91$. 
When the back-reaction cannot be ignored, the maximum eccentricity is not achieved at $c_{\rm io}=0$ ($I_{io,{\rm init}}=90^{\circ}$) but instead at (noting that Eq. (\ref{eq:c_io_vs_j_min}) describes a quadratic function with respect to $c_{io}$)
\begin{align}
    c_{io, {\rm lim}} = \frac{\eta}{2}\left(\frac{4}{5} j_{\rm lim}^2 - 1\right) \simeq - \frac{\eta}{2} \simeq -0.026 \left(\frac{a}{0.2\,{\rm AU}}\right)^{1/2}.
    \label{eq:c_io_lim}
\end{align}
In other words, the limiting eccentricity is achieved when the initial inclination between the inner and outer orbits is $I_{io, {\rm init}}\simeq 91.5^\circ$. Consequently, an initially retrograde orbit in general experiences a stronger LK excitation than a prograde one does.

Our prescription for tidal dissipation follows the static-tide model introduced by \citet{Hut:81}. The dissipation is parameterized through a constant time lag $t_{\rm lag}$ that is inversely proportional to the tidal quality factor $Q$. The characteristic decay timescale is \citep{Vick:19}
\begin{align}
    &t_{t}^{-1} =  \left(\Big{|}\frac{\dot{a}}{a}\Big{|}\sqrt{1-e^2}\right)_{e_{\rm max}}\nonumber \\
    =&6 k_{2p} t_{\rm lag} \frac{M_\ast}{M_p}\left(\frac{R_p}{a}\right)^5 \frac{n^2}{(1-e^2)^7}\left[f_1(e) - \frac{f_2^2(e)}{f_5(e)}\right],
\end{align}
where $k_{2p}$ is the planetary Love number and $n=2\pi/P$ is the mean motion; $f_{1,2,5}(e)$ are functions of eccentricity defined in \citet{Hut:81}.
An additional factor of $\sqrt{1-e_{\rm max}^2}$ is included in the timescale estimation because the time the planet spends at  $e\simeq e_{\rm max}$ is only a fraction of $\sqrt{1-e_{\rm max}^2}$ of the LK oscillation period.
We consider here only the planetary dissipation; the stellar dissipation is smaller by a factor of $(M_\ast/M_p)^2 (R_p/R_\ast)^5\sim 3\times10^3$ the values for $k_2$ and $t_{\rm lag}$ are similar. 
We assume $t_{\rm lag}=10\,{\rm s}$ as the default value in our simulations. Such a choice is motivated by the quality factor of our ice giants $Q\simeq 10^4$ \citep[see recent results on Uranus, ][]{Gomes2024}, while for Jupiter-mass planets have $Q=10^5 - 10^6$ (\citealt{Wu:05, Millholland:19} and references therein). 
To produce the vigorous internal heating of WASP-107 b with an internal temperature of $345\,{\rm K}$ \citep{Welbanks:24}, or equivalently a tidal heating rate of $L_t\simeq 10^{-7}\,L_\odot$, it also require $t_{\rm lag}\simeq 10\,{\rm s}$ (see later in Eq.~\ref{eq:L_t}).\footnote{As a caveat, this estimation assumes that all the internal heating is caused by the instantaneous tidal heating. The mean tidal heat rate averaged over the migration can be estimated by dividing the current orbital binding energy by the age of the system, leading to $\langle L_t \rangle \simeq 2\times10^{-7} L_\odot$. If the migration time is shorter, the mean heating rate can be even higher. The residual of this may also contribute to the observed heating, and we leave this possibility to future investigations. 
}
Correspondingly, the time lag for Jupiter is $t_{\rm lag}\simeq 0.1\,{\rm s}$, and \citet{Anderson:16} adopted $t_{\rm lag}=1\,{\rm s}$ during their high-e migration simulations. For completeness, we will also consider the case where $t_{\rm lag}=1\,{\rm s}$ in Sec.~\ref{sec:pop_tidal_diss}.

A window for successful migration within a given time can be defined once the tidal lag time $t_{\rm lag}$ is specified. One can equate the migration timescale and the age of the system (which we assume to be 600\,Myr for WASP-107 b-like systems). This leads to a requirement on the pericenter separation \citep{Vick:19}
\begin{align}
    r_p&\lesssim r_{p, {\rm mig}} = 0.040\,{\rm AU} \left(\frac{M_\ast}{0.69\,M_\odot}\right)^{2/7} \left(\frac{a_{\rm init}}{0.2\,{\rm AU}}\right)^{-1/7} \nonumber \\
    &\times \left(\frac{M_p}{0.12\,{M_j}}\right)^{-1/7} \left(\frac{R_p}{0.94\,R_j}\right)^{5/7} \left(\frac{k_{2p}}{0.37}\right)^{1/7} \nonumber \\
    &\times \left(\frac{t_t}{600\,{\rm Myr}}\right)^{1/7} \left(\frac{t_{\rm lag}}{10\,{\rm s}}\right)^{1/7},
    \label{eq:r_p_mig}
\end{align}
for the planet to be able to migrate within its age, where $r_{p, {\rm mig}}$ denotes the critical value of $r_p$ for the planet to migrate. 
Under quaLK, the pericenter separation further translates to a window on the required initial inclination between the inner and outer orbit, $I_{io, {\rm init}}$. As Eq.~(\ref{eq:c_io_vs_j_min}) is quadratic in $c_{io}$, its roots can be easily computed from 
\begin{equation}
    c_{io,{\rm mig}} = \frac{-\mathcal{B} \pm \sqrt{\mathcal{B}^2 - 4 \mathcal{A}\mathcal{C}}}{2 \mathcal{A} },
    \label{eq:c_io_roots}
\end{equation}
where
\begin{align*}
    \mathcal{A} &= \frac{15}{8}\frac{e_{\rm max}^2}{j_{\rm min}^2}, \quad \mathcal{B} = \frac{3}{8}\frac{e_{\rm max}^2}{j_{\rm min}^2} \eta \left(5 - 4 j_{\rm min}^2\right), \\
    \mathcal{C} &= \frac{3}{8}\frac{e_{\rm max}^2}{j_{\rm min}^2}\left[\frac{5}{4}\eta^2 - (3 + \frac{9}{4}\eta^2) j_{\rm min}^2 + \eta^2 j_{\rm min}^4 \right]  \\
    &+\epsilon_{\rm GR}\left(\frac{1}{j_{\rm min}} - 1\right)  
    + \frac{\epsilon_{\rm ST}}{15} \left(\frac{1 + 3 e_{\rm max}^2 + \frac{3}{8}e_{\rm max}^4}{j_{\rm max}^9} - 1\right),         
\end{align*}
with 
\begin{equation*}
    e_{\rm max}^2 = 1- j_{\rm min}^2 \equiv \left( 1 - \frac{r_{ p,{\rm mig}}}{a_{\rm init}}\right)^2.
\end{equation*}
We emphasize that the estimation above applies only to the quaLK systems. 
The relative importance of the octupole term can be quantified by the parameter \citep{Liu:15}
\begin{align}
    &\epsilon_{\rm oct} = \frac{M_\ast - M_p}{M_\ast + M_p} \frac{a}{a_o (1-e_o^2)} e_o \nonumber \\
    &=0.033 \left(\frac{M_\ast - M_p}{M_\ast + M_p}\right) \left(\frac{a}{0.2\,{\rm AU}}\right) \left(\frac{a_o(1-e_o^2)}{1.7\,{\rm AU}}\right)^{-1}\left(\frac{e_o}{0.28}\right),
\end{align}
which can be significant for the WASP-107 system.
As we will see later in Sec.~\ref{sec:mutual_inclination} (and consistent with previous analyses, e.g., \citealt{Liu:15, Petrovich:15, Naoz:16}), the octLK enables a much wider merger window in general with the addition of the octupole potential.

Throughout our simulations, we make a few simplifying assumptions: the inner planet's spin is instantaneously pseudo-synchronized~\citep{Hut:81} and aligned with the inner orbit. We fixed the stellar rotation period at the currently measured $P_{\ast}= 17.2\,{\rm days}$ \citep{Dai_2017}.
This relatively long period suggests that the precession rate induced by stellar rotation is typically 20 times smaller than that due to general relativity (see, e.g., eqs. A8 and A10 of \citealt{Anderson:16}). 
Moreover, the characteristic spin-down timescale (denoted with a subscript ``sd'') for the host star due to magnetic breaking is \citep{Barker:09, Anderson:16}
\begin{equation}
    \tau_{\rm sd} = \frac{1}{\alpha \Omega_\ast^2} \simeq 3.7\,{\rm Gyr}\left(\frac{\alpha }{1.5 \times 10^{-14}\, {\rm yr}}\right)^{-1}\left(\frac{P_\ast}{17.2\,{\rm d}} \right)^2.
\end{equation}
The fiducial value of $\alpha$ is taken from \citet{Barker:09} for K stars (see also, \citealt{Curtis}). This is much greater than the migration timescale and hence the stellar spindown is insignificant. We separately verified this through a set of numerical simulations including the spindown which shows little difference from the main results we present.  
The direction of stellar spin vector $\vect{S}_\ast$, however, precesses due to its interaction with the orbital angular momentum (see appendix A2 of \citealt{Anderson:16}). Following the precession of the stellar rotation is important for calculating the stellar obliquity which is an observed quantity for WASP-107 b \citep{Rubenzahl}.


\subsection{Tidally driven mass loss and disruption}
\label{sec:mass_loss}

During a high-e migration, the close pericenter passage of the planet makes it venture to tidal disruption by the host star. 
Motivated by numerical simulations performed by \citet{Guillochon:11}, \citet{Anderson:16} used a hard boundary to capture the disruption. Whenever the pericenter separation $r_p$ is less than $2.7 r_t$ ($r_t\equiv R_p (M_\ast/M_p)^{1/3}$ is the tidal radius), the migrating planet is considered to be instantaneously and fully disrupted.
We include a more flexible tidally-driven mass loss prescription than that adopted by \citet{Anderson:16}. 
A key question we would like to address is whether WASP-107 b could have started as a much more massive planet (say Jupiter-mass), and was subsequently reduced to its observed Neptune-like mass during the postulated high-e migration by \emph{partial} disruption. \citet{Guillochon:11} has performed extensive 3-D hydrodynamical simulations of multi-orbit encounters of a migrating hot Jupiter with the host star with pericenter close to a few times the tidal radius. \citet{Guillochon:11} found that the mass loss in each close encounter is roughly exponentially related to the pericenter distance normalized by the tidal radius:
\begin{equation}
    \Delta M_p^{\rm (1\, pass)}/M_p = -A {\rm Exp}[ - B (r_p/r_t)] < 0,
    \label{eq:dMp_1_pass}
\end{equation}
where $A$ and $B$ are coefficients of a phenomenological fit to their hydrodynamic simulations of mass loss (see their fig. 10). 
We stress that this mode of mass loss is primarily due to the quick growth and breakup of the dynamical tides on the planet at $r_p>2r_t$ \citealt{Mardling:95, Ivanov:04, Ivanov:07, Wu:18, Vick:18, Yu:21, Yu:22}) rather than direct Roche lobe overflow (which happens at $r_p\approx 2 r_t$ when $M_p\ll M_\ast$; \citealt{Eggleton:83}).
While the mass loss at each close encounter can be highly stochastic, we use Eq.~(\ref{eq:dMp_1_pass}) to approximate the averaged effect. 
The value of $B$ is set to $B=9.25$, corresponding to the sharp dependence of the dynamical tide on the pericenter distance. The slope is consistent with the solid-orange line in fig. 10 in \citet{Guillochon:11}. We leave the parameter $A$ as a free variable to be explored during the simulations. It is sampled log-uniformly from the range $A\sim 10^3-10^5$, corresponding to a $10^{-8} - 10^{-6}$ fractional mass loss per pericenter passage. As we will see later in Fig.~\ref{fig:outcome_vs_mass_loss},  $A$ effectively controls the boundary between disruption and survival of the migrating planet.  The range we assume reproduces the survival threshold found in \citet{Guillochon:11} while allowing partial disruptions of the planet. 
While most of the mass loss occurs near periastron passage, we can estimate the orbital averaged rate as
\begin{equation}
    \dot{M_p} = \frac{\Delta M_p^{\rm (1\, pass)}}{P}, 
    \label{eq:dMp_mean}
\end{equation}
where $P$ is the orbital period. 
For simplicity, we fix the planetary radius when its mass changes, an assumption appropriate if the planet's equation of state can be approximated by a $p\propto \rho^2$ polytrope.

We further note that both the tidal radius $r_t$ and the pericenter distance $r_p$ may evolve as a result of the mass loss in the previous passage.  The tidal radius $r_t$ changes because the density of the planet adjusts after the mass loss. If the density of the planet decreases, the tidal radius $r_t$ increases (the surface gravity of planet is weaker and the planet is more loosely bound). The next encounter will lead to more intensive mass loss. As such, the planet may enter a run-away mass loss, i.e. disruption. 

Mass loss may change the orbit of the planet as well. Here we consider two limiting cases. The first limiting case is the so-called conservative mass transfer frequently used when one object in a binary overfills its Roche Lobe \citep{Sepinsky:07}. Here the host star and WASP-107 b are treated like a binary system, whose total mass and total orbital angular momentum are conserved. If so, we have~\citep{Sepinsky:07}
\begin{align}
    &\left(\frac{\dot{a}}{a}\right)_{\rm MT, con} = 2 \frac{1+e}{1-e}\left(\frac{M_p}{M_\ast} - 1\right) \frac{\dot{M_p}}{M_p},\label{eq:da_MT} \\
    &\dot{e}_{\rm MT, con} = (1-e) \left(\frac{\dot{a}}{a}\right)_{\rm MT, con}. 
\end{align}
The equations above also assume that mass transfer happens instantaneously at the pericenter, and we ignore finite-size effects.  \citet{Sepinsky:07} suggest that the mass transfer creates only a torque without a radial acceleration [see their eqs. (B5), (B6), and (16-18)]. Consequently, it does not lead to pericenter precession and we have $\dvect{e}_{\rm MT, con} = \dot{e}_{\rm MT, con}  \uvect{e}$. We will refer to this scenario as ``conservative mass transfer'' with subscript ``con''. 

However, \citet{Guillochon:11} suggest that the host star may accrete only a small fraction of the mass loss from the planet.  
Furthermore, it is unclear if the mass accreted by the host star can quickly transfer the angular momentum it carries back to the orbit of the planet. As another limiting case, we assume the orbit loses all of the angular momentum carried by the mass loss:
\begin{equation}
    \left(\frac{\dot{J}}{J}\right)_{\rm MT, non-con} = \frac{\dot{M}_p}{M_p}. 
\end{equation}
Because $M_\ast \gg M_p$, we simply set $\dot{a}_{\rm MT, non-con}=\dot{e}_{\rm MT, non-con}=0$. We will refer to this prescription as ``non-conservative mass loss''. We note that neither prescription is perfect, the reality is most likely somewhere in between. Yet given the nature of simulations done by \citet{Guillochon:11}, the non-conservative scenario may be more favored. As we will see later in Fig.~\ref{fig:formation_rate}, the formation of WASP-107 b-like planet also favors the non-conservative limit. 

Similar to the migration window defined in Eqs.~(\ref{eq:r_p_mig}) and (\ref{eq:c_io_roots}), the requirement for the planet to survive disruption also constrains the allowed range on $r_p$ and further $I_{io, {\rm init}}$. As we will see later in Fig.~\ref{fig:outcome_vs_mass_loss}, the requirement can be written as 
\begin{equation}
    r_p \gtrsim r_{p, {\rm dis}}\equiv c_{\rm dis} r_t,
    \label{eq:r_p_dis}
\end{equation}
where $c_{\rm dis}$ is a numerical constant ($c_{\rm dis}\simeq 2.7$ based on \citealt{Guillochon:11} and our choice of $A$ in Eq.~(\ref{eq:dMp_1_pass}) explores the range $c_{\rm dis} \in [2, 3]$). In terms of $I_{io, {\rm init}}$, the window can be obtained using again the quadratic formula in Eq.~(\ref{eq:c_io_roots}) but this time with $e_{\rm max} = 1-r_{p,{\rm dis}}/a_{\rm init}$. Therefore, a successful migration requires the pericenter separation satisfies $r_{p, {\rm dis}} \lesssim  r_p \lesssim r_{p, {\rm mig}}$.

\subsection{Radius Inflation}
\label{sec:R_infl}

The radius of WASP-107 b is inflated. The planet has a Jupiter-like radius of 0.948$\pm0.03\, R_j$ \citep{Dai_2017} while its mass is only 10\% of that of Jupiter: $30.5\pm1.7M_\oplus$ \citep{Piaulet}. The mean density $\sim$0.1 g~cm$^{-3}$ is a whole order of magnitude lower than Jupiter's. 
To address the possibility of radius inflation,
we incorporate a prescription following \citet{Thorngren:21} for a subset of our simulations. The equilibrium radius of the planet is given by:
\begin{align}
    R_{\rm eq} =& 1.21 R_j \left(\frac{M_p}{M_{j}}\right)^{-0.045} \nonumber \\
    &\times \left(\frac{F}{\rm 10^9erg\,s^{-1}\,cm^{-2}}\right)^{0.149-0.072\lg (M_p/M_{j})},
    \label{eq:Req}
\end{align}
with the effective flux $F$ given by 
\begin{align}
    F = F_{\rm irr} + F_t = \frac{L_\ast}{4 \pi r_{\rm eff}^2} + \frac{L_t}{4\pi R_p^2}.
\end{align}
We have $1/r_{\rm eff}^2 = 1/(a^2\sqrt{1-e^2})$, which preserves the insolation averaged over each orbit. We have also added the tidal heating, 
\begin{equation}
    L_{t} = 3 k_{2p}  t_{\rm lag} n^2 \left(\frac{G M_\ast^2}{R_p}\right)\left(\frac{R_p}{a}\right)^6 \frac{\left[f_1(e) - \frac{f_2^2(e)}{f_5(e)}\right]}{(1-e^2)^{15/2}},
    \label{eq:L_t}
\end{equation}
assuming that it is evenly distributed throughout the planet. 
We note that $R_{eq}=0$ when $F=0$ in this simple prescription. Therefore, we add a lower limit on the equilibrium radius using the mass of the planet and empirical mass-radius relationship \citep{Chen_Kipping}. When the instantaneous planetary radius deviates from the equilibrium radius, it evolves over time according to 
\begin{align}
    &\frac{dR}{dt} = \frac{R_{\rm eq} - R}{\tau}, \\ 
    &\tau = 
    \begin{cases}
    \tau_{\rm def}, \quad R>R_{\rm eq},\\
    \tau_{\rm inf}, \quad R<R_{\rm eq}.
    \end{cases}
    \label{eq:tau_d_r}
\end{align}
The timescales $\tau_{\rm def}$ and $\tau_{\rm inf}$ separately control the deflation and inflation timescale of the planet. Deflation timescale is most likely comparable to or longer than the duration of our simulation (\citealt{Thorngren:21} found evidence for delayed cooling, and they used $\tau_{\rm def}=$ 0.5 Gyr). We similarly adopted $\tau_{\rm def}=0.5\, {\rm Gyr}$. On the other, the inflation timescale is much more critical in our study (\citealt{Thorngren:21} found evidence for rapid reinflation of hot Jupiters). We thus uniformly sampled $\lg \left(\tau_{\rm inf}/{\rm Myr}\right)$ between 0 to 2.5 when the planetary radius is allowed to change. We do not consider greater values of $\tau_{\rm inf}$ as it changes the radius by little. This is equivalent to turning off radius evolution which we do for another subset of our simulations. 
Furthermore, we find our radius inflation prescription affects only the final planetary radius but has weak impacts on the other properties of the system. The reason is that a significant radius inflation can happen only near the end of the migration where the inner binary and the perturbing planet have decoupled. We will discuss this in more details in Sec.~\ref{sec:results_examples} when we examine specific examples. 
As a caveat, we note that the results of \citet{Thorngren:21} were calibrated only for planets with $M_p>0.5\,M_{j}$. We have extrapolated their results to $M_{j}\in (0.1,\ 0.5)\,M_j$ which is the regime we are studying here for WASP-107 b. We have compared the results with and without radius inflation, and the qualitative conclusions of this paper remain consistent. Future work with a more elaborate radius inflation model in the $M_j\in (0.1,\ 0.5)\,M_j$ regime should help improve our results.


\section{Case Study}
\label{sec:results_examples}

We performed a population synthesis study to explore how different initial conditions and system parameters determine the formation of WASP-107b-like systems. The parameters we sampled and their ranges are shown in Table~\ref{tab:pop_syn_par}. Before talking about the population-level outcome of our simulations, let us first examine some specific examples.

\begin{figure}
    \centering
    \includegraphics[width=0.9\linewidth]{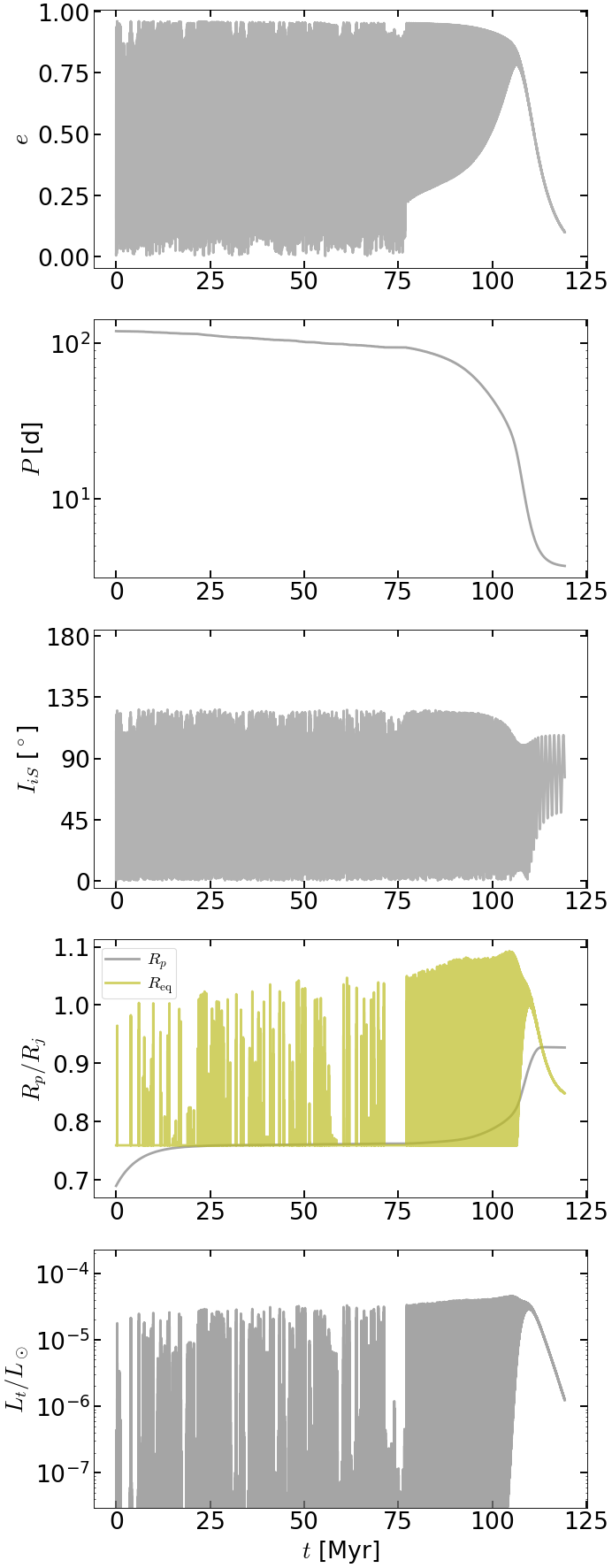}
    \caption{Evolution trajectory of a potential WASP-107b progenitor. From top to bottom, we show the orbital eccentricity, orbital period, stellar obliquity, planetary radius, and tidal heating luminosity. This high-e migration reproduces multiple observable features of WASP-107 b: it successfully migrated from an initial orbital period $>100$ days to an orbital period $<5$ days in $\sim 100\,{\rm Myr}$ with an inflated radius and nearly polar obliquity.
    The current tidal heating can be as high as 10$^{-6}L_\odot$.
    }
    \label{fig:wasp_107b_sample_traj_2_2731}
\end{figure}

Fig.~\ref{fig:wasp_107b_sample_traj_2_2731} shows the evolution track for a particular simulation that reproduces multiple observed properties of WASP-107 b. From top to bottom, we show the evolution of the inner planet's orbital eccentricity $e$, orbital period $P$, the stellar obliquity, the planet radius $R_p$, and the luminosity due to eccentricity tides $L_t$. The system is evolved with octupole LK oscillations and nonconservative mass loss. The initial planet mass is $0.25\,M_j$ and stays nearly constant during the evolution. In other words, mass loss was minimal during the whole migration process. The inner planet has an initial orbit of $a_{\rm init}=0.42\,{\rm AU}$ and initial $e_{\rm init}=0.01$. The star's spin vector $\vect{S}_\ast$ is initially aligned with the inner orbital angular momentum vector $\vect{J}$. 
The outer orbit has $M_c=0.62\,M_j$, $e_{o, {\rm init}}=0.33$ and the initial mutual inclination between planet b and c is $I_{io,{\rm init}}=64^\circ$. 

The high-e migration can be broken into three stages. When $t<75\,{\rm Myr}$, the inner planet experiences eccentricity excitation through the LK cycle. Because of the octupole terms, the eccentricity excitation exhibits stochasticity \citep{Li:14b, Liu:15}. The pericenter $r_p$ occasionally reaches very close to the host star. 
Whenever this happens, the eccentricity tide is strongly amplified, leading to the spikes in the tidal luminosity and equilibrium planet radius in the last two panels of Fig. \ref{fig:wasp_107b_sample_traj_2_2731}. However, the radius of the inner planet stays more or less constant during this stage, because the planet is still far from the host star for most of the orbit. The orbit-averaged insolation and tidal heating are not strong enough to consistently inflate the planet. We assume an inflation timescale of $\tau_{\rm inf}=5.8\,{\rm Myr}$ in this particular simulation.

The second stage of the high-e migration is roughly between $75\,{\rm Myr}<t<110\,{\rm Myr}$. Here, the semi-major axis of the planet has decayed so much that short-range forces (including relativity, the conservative planetary tide, and rotation-induced quadrupoles in both the star and the planet) become more significant and quenched the octupole LK effects (see sec. 3.1 in \citealt{Anderson:16} for more detailed discussions; also \citealt{Fabrycky:07, Liu:15}). However, quadrupole LK effects remain important, and the eccentricity oscillation becomes regular with a gradually shrinking range. The pericenter distance of the planet follows a well-defined envelope, which is reflected in the envelope of the tidal heating rate. The orbital period of the planet declines steadily as shown in the second panel.

Finally, at around $t\simeq 110\,{\rm Myr}$, the orbital decay timescale becomes shorter than the LK oscillation timescale, and the inner planet decouples from the secular influence of the outer planet. The inner orbit evolves under the influence of eccentricity tides. The inner planet circularizes along a trajectory with nearly constant orbital angular momentum, so $P$ and $e$ decrease while the pericenter separation $r_p$ gradually increases. 

The peak of tidal heating and radius inflation happened between $100\,{\rm Myr} < t < 110\,{\rm Myr}$ when the orbital semi-major axis has decayed significantly, amplifying the insolation from the host star. Meanwhile, the orbit stays eccentric enough with the pericenter separation $r_p$ stays near its minimum, thereby boosting the tidal heating rate. It is the combination of intense irradiation and tidal heating that inflated the planet by about 20\% during this time from 0.76$R_{\rm j}$ to 0.93 $R_{\rm j}$ similar to the observed value.  
Inflation at this late stage is also crucial for the planet to survive the migration early on. 
The long deflation timescale $\tau_{\rm def}=500\,{\rm Myr}$ allows the planet to stay inflated even though tidal heating is lower at the very end of the simulation than during the peak of tidal heating between $100\,{\rm Myr} < t < 110\,{\rm Myr}$.

Meanwhile, the obliquity is attracted to around $90^\circ$ after decoupling from the perturber (third panel). This is due to the dynamical attractor discussed in \citet{Liu:18} and \citet{Yu:20}. The effective axis around which the stellar spin precesses changes from the outer orbital angular momentum during the LK cycle to the inner orbital angular momentum when the inner orbit decouples from the perturber, with the opening angle being nearly constant during the transition. This produces a final obliquity consistent with observations \citep{Dai_2017,Rubenzahl}. 

We terminate the simulation when the inner eccentricity reaches 0.1. The inner orbital period decays to 3.7 days. The tidal heating rate is smaller than its peak value by nearly two orders of magnitude yet can still be significant at a level of $10^{-6}\,L_\odot$.

\begin{figure}
    \centering
    \includegraphics[width=0.82\linewidth]{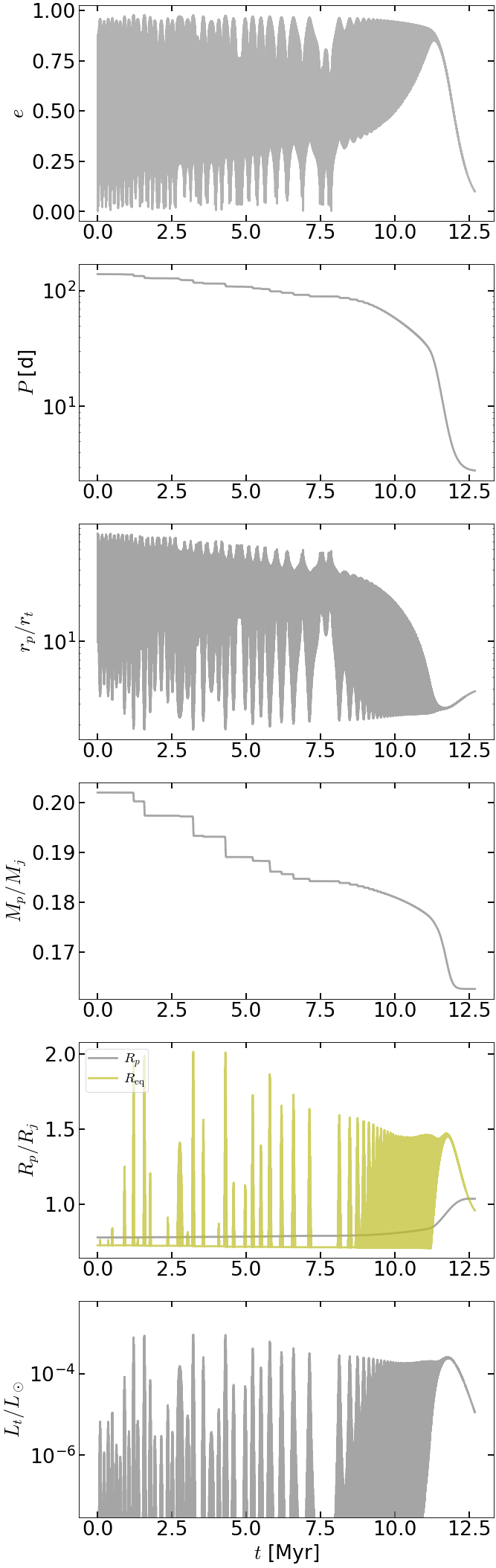}
    \caption{Similar to Fig.~\ref{fig:wasp_107b_sample_traj_2_2731} but for a system that experiences substantial mass loss (fourth panel). The planet has $M_{p0}=0.20\, M_j$ initially and loses $0.04\,M_j$ of its mass during the migration. This is one of the cases where WASP-107 b might lose significant mass. A stronger mass loss likely leads to disruption of the planet.}
    \label{fig:wasp_107b_sample_traj_9_994}
\end{figure}

A keen reader may ask what if the inner planet ventures closer to the host star during that initial stage of migration where octupole LK oscillation stochastically excites the orbital eccentricity. Could WASP-107 b lose significant mass at this stage? Yes, and we now present a case where WASP-107 b might lose substantial mass during its high-e migration in Fig.~\ref{fig:wasp_107b_sample_traj_9_994}. 
The initial conditions for this system are $(M_{p},\, R_{p},\, a,\, M_c,\, e_o,\, I_{io})_{\rm init} = (0.20 M_j,\, 0.78 R_j,\,0.46\,{\rm AU},\, 1.1\,M_j,\,0.20,\, 70^\circ)$.
A new panel showing the evolution of planetary mass $M_p$ is added to the figure. We also show the pericenter separation normalized by the tidal radius. Notice that the whole high-e migration happens on a much faster timescale here $\sim10$ Myr rather than the $\sim100$ Myr in the previous example. This is because as the pericenter distance of planet b  comes closer to the host star, the strong tides speed up the inward migration (see also Fig.~\ref{fig:t_mig} later). Also the strong tides unbind a fraction of the planet episodically whenever $r_p/r_t$ is small.  The orbit stays at such small $r_p$ only momentarily due to the stochastic nature of the octupole LK oscillation. The planet thus could avoid a runaway disruption (see first four panels of Fig. \ref{fig:wasp_107b_sample_traj_9_994}). 
Besides the initial LK cycles, mass loss can also happen in another epoch near the end of the migration (around $t=12\,{\rm Myr}$ in the example shown). The pericenter separation stays small while the orbital period decreases rapidly, which enhances the mass loss rate according to Eq.~(\ref{eq:dMp_mean}). Radius inflation may also happen at the same time, keeping $r_p/r_t$ small even if the physical $r_p$ increases as the binary circularizes.
The integrated mass loss over the entire high-e migration is about 20\% of the initial mass. We will show with population synthesis in the next Section that  $>20\%$ mass loss usually results in a run-away disruption of the planet. The short $\sim10$ Myr requires that the high-e migration was only recently initiated on WASP-107 b. 
Such a delayed migration may happen if planets b and c were only recently scattered into sufficiently inclined orbits; see, e.g., \citet{Lu:24}.


\section{Population Study}
\label{sec:results_pop_level}

\begin{table}
    \centering
    \begin{tabular}{ccc}
    \hline
    \hline
        Parameter & meaning &  range \\
    \hline
         $M_{p, {\rm init}}/ M_j$ & Initial inner planet mass & (0.1, 0.5)\\
         $R_{p, {\rm init}}/R_j$ & Initial planet radius & (0.6, 1.5)\\
         $a_{\rm init}/{\rm AU}$ & Initial inner semi-major axis & (0.15, 0.5)\\
         $M_c/M_j$ & Outer planet mass & (0.36, 1.5)\\
         $\cos I_{io, {\rm init}}$ & Initial i-o inclination & (-0.9, 0.75)\\
         $e_{o, {\rm init}}$ & Initial outer eccentricity & (0.2, 0.45)\footnote{We sample $e_{o0}$ only when the octupole LK is used. For quadrupole LK, we fix $e_o=0.28$. }\\
         $\phi_{o, {\rm init}}$ & Outer argument of periastron & (0, $2\pi$)\\
         $\lg A$ & Tidal mass loss rate, see Eq.~(\ref{eq:dMp_1_pass}) & (2.8, 5)\\
         $\lg \tau_{\rm inf}$ & Inflation timescale in Myr (Eq.~\ref{eq:tau_d_r}) & (0, 2.5)\\
         \hline
    \end{tabular}
    \caption{Parameters we sample for the population synthesis study. All of them are sampled uniformly within the specified range. }
    \label{tab:pop_syn_par}
\end{table}

We run about $2\times 10^5$ simulations to understand the probability of the high-e migration scenarios for the WASP-107 system, and how the initial system parameters translate to the testable observables at the end of the simulation. 
Table \ref{tab:pop_syn_par} summarizes the relevant initial parameters and their prior region. Including radius inflation or not does not impact significantly the formation rate and other population properties, because inflation happens only after the inner binary decouples from the outer planet. Therefore, in the results below we focus on the case where $R_p$ is kept constant throughout the evolution. 
We note that within quaLK, one can show that
\begin{equation}
    \vect{e}_o \cdot \left(\frac{d \vect{e}_o}{dt}\right)_{\rm quaLK} = 0.
\end{equation}
Therefore, for quaLK runs, one can fix the eccentricity of the outer planet at its currently observed value $e_o=0.28$ \citep{Piaulet}. After including octupole LK terms, however, $e_o$ is no longer a constant and we sample its initial value uniformly between 0.2-0.45 near the measured value of 0.28$\pm$0.07 \citep{Piaulet}. The outer eccentricity typically decreases during migration as the LK oscillation transfers angular momentum from the inner orbit to the outer one. Therefore, we do not consider cases where $e_{o, {\rm init}}<0.2$.

We set the coordinates such that the initial angular momentum vector of the outer planet $\uvect{J}_{o,{\rm init}}$ is along the $\uvect{z}$ axis, and the Cartesian components of the inner angular momentum are $\uvect{J}_{\rm init} = (\sin \iota_{io, {\rm init}}, 0, \cos \iota_{io, {\rm init}})$ where $\iota_{io, {\rm init}}$ is the initial mutual inclination between the inner and outer orbits. The spin vector of the host star is assumed to be aligned with the inner planet's orbital axis, $\uvect{S}_{\ast, {\rm init}} = \uvect{J}_{\rm init}$. We will denote the obliquity as $I_{iS}$ with $\cos I_{iS} = \uvect{J} \cdot \uvect{S}_\ast$. 
\citet{Vick:23} suggests that the stellar obliquity can be excited by the protoplanetary disk before the onset of the LK oscillation. We defer investigations of this possibility to future studies. 
We initialize the outer eccentricity as $\vect{e}_{o, {\rm init}} = e_{o, {\rm init}}(-\sin \phi_{o, {\rm init}}, \cos \phi_{o, {\rm init}}, 0)$ where the argument of periastron $\phi_{o, {\rm init}}$ is uniform between 0 to $2\pi$. The initial inner eccentricity is set to $(0, 0.01, 0)$ for all systems. 

Once the initial conditions are set, we evolve each system for $600\,{\rm Myr}$ or if one of the termination conditions is met: 1) if the planet's mass decreases below $0.03 M_j$ or its orbit becomes unbound [as $a$ increases due to mass transfer; Eq.~(\ref{eq:da_MT})]. We flag the system as being disrupted. 2) If the inner orbital period decreases below $10\,{\rm d}$ \emph{and} the inner eccentricity is below $0.1$, we then flag it as a successful migration.  3) If none of the termination flags are met at the end of the $600\,{\rm Myr}$ integration, the system is labeled as `no or partial migration'. 4) We also check for orbit crossing in post-processing. If $a (1+e) >= a_o(1-e_o)$ at any moment during the evolution, we assume such systems cannot maintain long-term stability and flag them also under the disruption category.  

\begin{figure}
    \centering
    \includegraphics[width=1\linewidth]{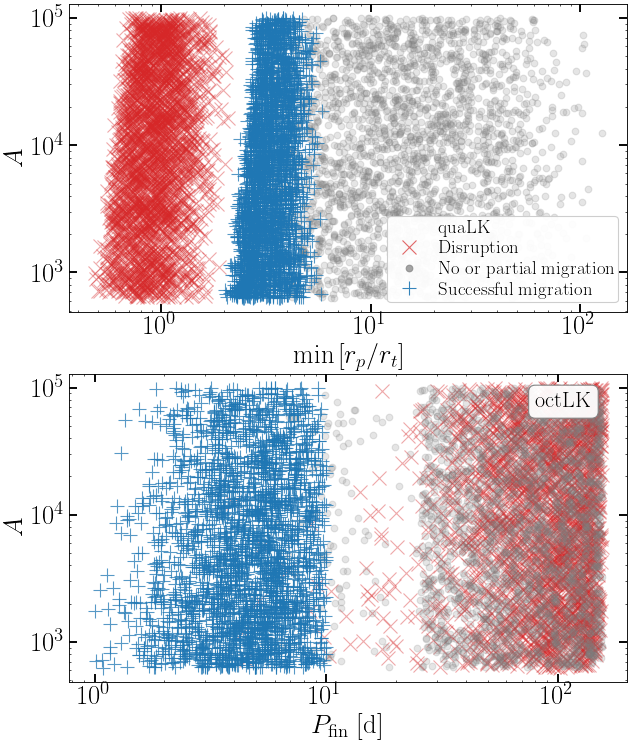}
    \caption{Population synthesis outcome over the parameter space of the phenomenological parameters $A$  governing the mass loss [Eq.~(\ref{eq:dMp_1_pass})] as well as ${\rm min} (r_p/r_t)$ (top panel) and $P_{\rm fin}$ (bottom panel). The top panel is evolved under quaLK and the bottom panel also includes octupole interactions, yet the overall result stays the same regardless of the LK prescription. All panels assume the mass loss is nonconservative. 
    The value of $A$ effectively determines the boundary between disruption and survival, with increasing $A$ shifting the critical periastron separation $r_p$ for successful migration and the final orbital period of migrated systems both to greater values. }
    \label{fig:outcome_vs_mass_loss}
\end{figure}

The simulation outcome is summarized in Fig.~\ref{fig:outcome_vs_mass_loss}. In the top panel, we see that varying the phenomenological parameter $A$ in Eq.~(\ref{eq:dMp_1_pass}) effectively changes the boundary between disruption and successful migration from ${\rm min}[r_p/r_t]=2$ to 3, covering the value of 2.7 suggested in \citet{Guillochon:11}. We discuss its implication of the planetary mass loss in Sec.~\ref{sec:pop_mass_loss}.   
In the bottom panel, we further show the outcome as a function of the final orbital period. As we have seen for the case studies in the previous section, high-e migration can be viewed as a two-step process where in the first step the LK cycle periodically extracts angular momentum from the inner orbit while keeping its energy largely intact, and in the second step tidal dissipation decouples the inner planet from the outer planet and circularizes the inner orbit along a trajectory with nearly conserved orbital angular momentum. Consequently, the final semi-major axis when the orbit is circularized is about twice the minimum pericenter separation when the orbit was highly eccentric as $a(1-e^2)\simeq {\rm const}$. Notice many tidally disrupted planets do not complete the first step before disruption, and their `final' orbital periods, reported at the time of disruption, are high and approximately equal to the initial periods.

\subsection{WASP-107 b probably did not lose $>20$\% of its initial mass}
\label{sec:pop_mass_loss}

\begin{figure}
    \centering
    \includegraphics[width=1\linewidth]{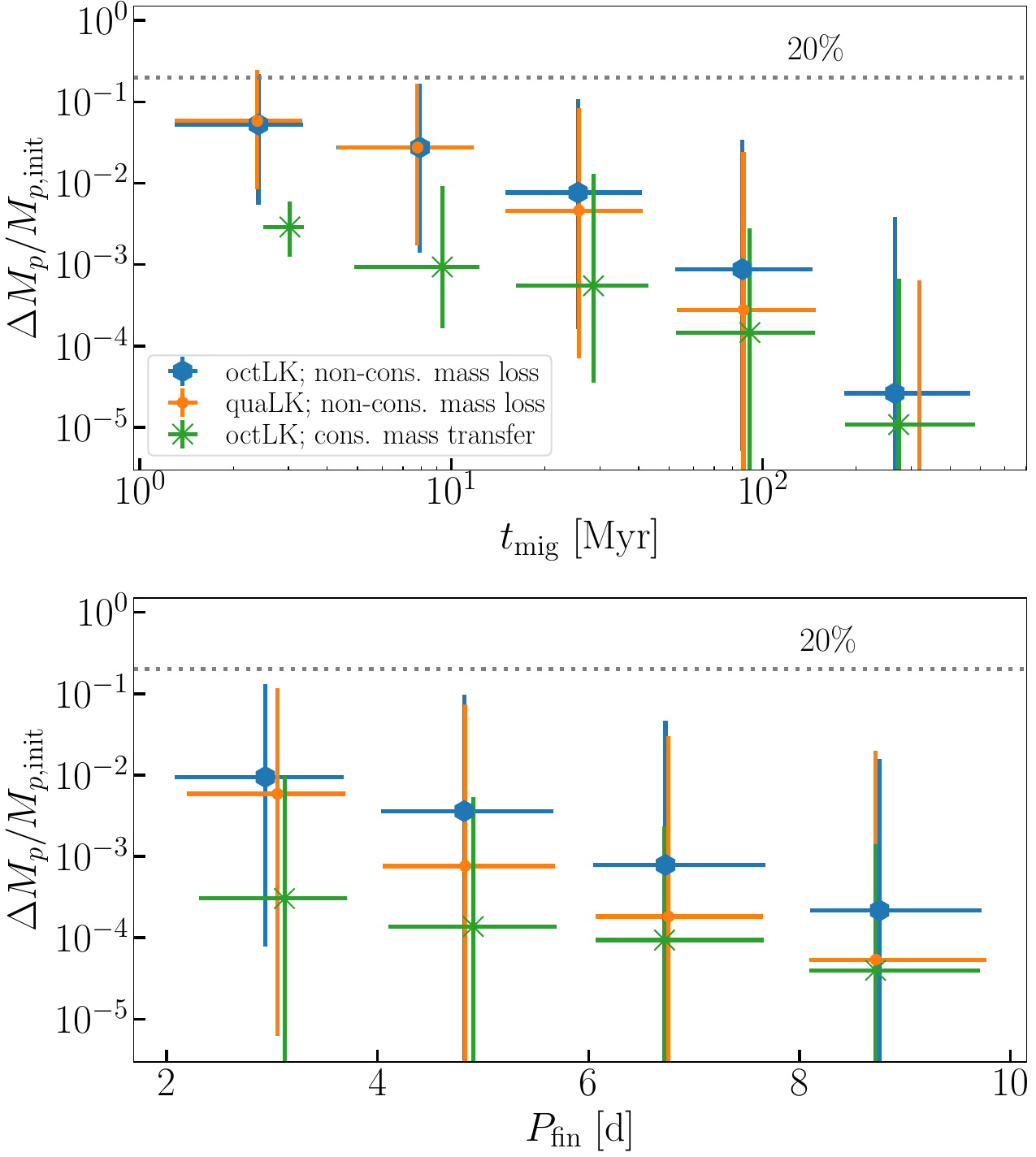}
    \caption{Fractional mass loss of successfully migrated planets as functions of the migration time (top) and the final orbital period (bottom). Error bars represent the 10th and 90th percentiles of the distribution within each bin. 
    The planet can lose up to $\sim 20\%$ of its initial mass while surviving partial disruption by the host star if the mass loss is non-conservative. A significant mass loss can happen if a system migrates with 10s of Myrs, as such a system can have small $r_p/r_t$ during migration. The correlation between $r_p$ and $P_{\rm fin}$ sees large scattering, so a system with $P_{\rm fin}\simeq 5\,{\rm d}$ can still experience up to $\sim 10\%$ mass loss. }
    \label{fig:dMp}
\end{figure}

We explored a wide range of mass loss efficiency due to dynamical tides by varying $A$ between $10^{2.8}$ to $10^5$ in the phenomenological mass loss model Eq.~(\ref{eq:dMp_1_pass}). However, our population synthesis suggests that WASP-107 b did not lose more than 20\% of its mass during the postulated high-e migration. In other words, it did not start as a Jupiter-mass object. The reason is as follows.

Tidal interaction has a very steep dependence on the pericenter distance, and we empirically find only systems with $r_p/r_t\lesssim 3$ may experience significant mass loss. 
This can be understood by examining the excitation of the planetary dynamical tide \citep{Press:77} that leads to the mass loss in \citet{Guillochon:11}. The characteristic energy acquired by the dynamical tide per pericenter passage can be written as (e.g., \citealt{Yu:21})
\begin{align}
    \frac{\Delta E_{p,{\rm DT}}}{E_p} \simeq 5 \pi k_{2p}\left(\frac{r_t}{r_p}\right)^6 |K_{22}|^2,
\end{align}
where $E_p\equiv G M_p^2/R_p$. Besides the spatial overlap $\sim (r_t/r_p)^6$, there is also a temporal overlap between the prograde planetary f-mode (with spherical harmonic $l=m=2$) and the orbital drive, $K_{22}$, that enters the expression, and it is evaluated as \citep{Lai:97}
\begin{align}
    |K_{22}| \simeq \frac{\sqrt{2} z^{5/2}e^{-2z/3}}{\sqrt{15}}\left(1-\frac{\sqrt{\pi}}{4\sqrt{z}}\right),
    \label{eq:K_lm_para}
\end{align}
where $z\equiv \sqrt{2} \omega/\Omega_{\rm peri} (\sim 10$ for systems considered in this study), with $\omega \simeq \sqrt{G M_p/R_p^3}$ is the planetary f-mode frequency (which we have approximated by the planetary dynamical frequency) and $\Omega_{\rm peri}=\sqrt{G(M_\ast + M_p)/r_p^3}$. The $|K_{22}|$ factor suppresses the tidal energy exponentially with increasing $r_p$, which, together with the work \citet{Guillochon:11}, motivates our Eq.~(\ref{eq:dMp_1_pass}).\footnote{Note, however, that we do not explicitly track the dynamical tide in our population synthesis as it is history-dependent and needs to be tracked on an orbit-by-orbit basis which is too expensive to follow (but see \citealt{Vick:19}).}

On the other hand, once the mass loss is initiated, it can lead to a run-away process. The tidal radius increases as the planet loses mass because $r_t \propto M_p^{-1/3}$ whereas $R_p$ stays nearly constant. Therefore, $r_p/r_t$ effectively decreases after the mass loss in a previous pericenter passage. Second, a less massive inner planet is more susceptible to eccentricity excitation from the LK cycle. Both effects increase the ratio of $r_p/r_t$, causing a further increase in mass loss, disrupting the planet in a run-away process. The run-away is also demonstrated by the formation of a gap between the successfully migrated population and those disrupted by the host star as shown in the top panel of Fig. \ref{fig:outcome_vs_mass_loss}. Any planet within the gap will quickly evolve to join the disrupted population.

While changing $A$ affects the boundary, the fractional mass loss does not sensitively depend on the value of $A$. Instead, we show in Fig.~\ref{fig:dMp} that it depends mostly on the migration time and the final orbital period, both of which further track $r_p$ during the migration. The mass loss is limited to $<20\%$ with a smaller $r_p$ (that is, shorter migration time and smaller $P_{\rm fin}$) leading to a greater fractional loss. The choice of quaLK or octLK does not affect the conclusion. On the other hand, a mass loss exceeding $1\%$ requires it to be largely non-conservative. Because mass loss happens when the orbit is highly eccentric with $(1-e)\ll 1$, any conservative mass transfer exceeding 1\% of the initial mass can disrupt the orbit due to the $(1-e)$ factor in the denominator in Eq.~(\ref{eq:da_MT}).


The relative formation rate of WASP-107 b also depends on the mean density of the planet $\bar{\rho}_p$ as shown in the middle panel of Fig. \ref{fig:formation_rate}. As the boundary between survival and disruption is a sensitive function of $r_p/r_t$, and $r_t\propto \rho_p^{-1/3}$, the formation rate thus favors planets with a high density. Empirically, the formation rate is proportional to $\lg \bar{\rho}_p$.  A higher density planet is more resistant to rapid mass loss and disruption. We thus conclude that the initial mass of WASP-107b is likely below $0.17\,M_j$ and it lost less than 20\% of its initial mass during the migration. As a caveat, however, our model does not constrain mass loss after the planet has finished high-e migration due to say photoevaporation \citep{Spake,Allart,Kirk,Spake2021}. \citet{Wang_Dai} suggests that the mass loss rate is currently low enough $1M_\oplus$/Gyr but could have been faster when the star was more active.



\subsection{Initial Mutual Inclination between WASP-107 b and c} 
\label{sec:mutual_inclination}

\begin{figure}
    \centering
    \includegraphics[width=1\linewidth]{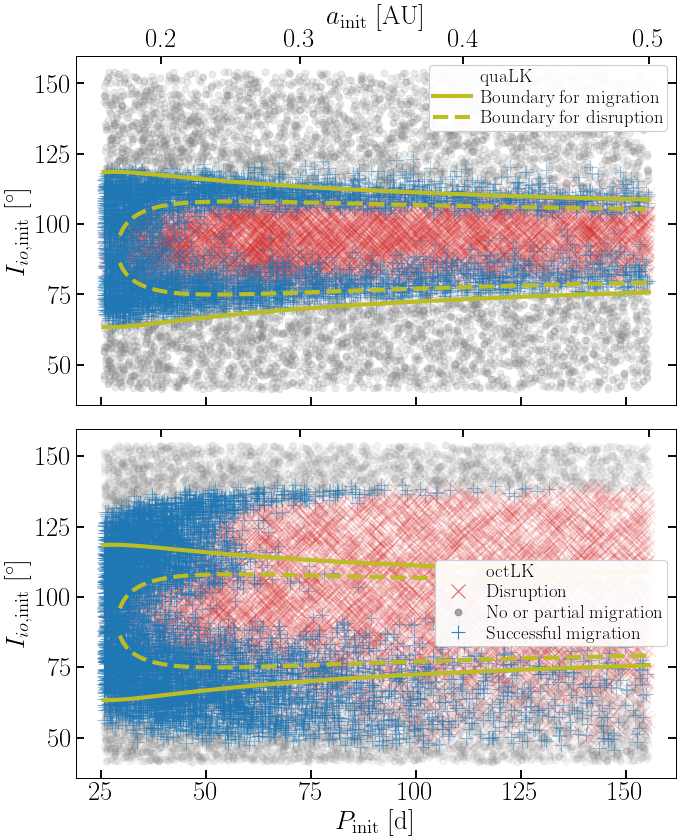}
    \caption{Outcome vs. initial inclination and period. The top panel shows systems evolved under quaLK and the bottom those under octLK. Non-conservative mass loss is assumed for both panels. 
    For systems with $P_{\rm init}\gtrsim 75\,{\rm d}$, including the octupole effects make a significant difference. Under quaLK, the outcome is almost symmetric about $\simeq 92^\circ$ and the successfully migrated systems concentrated in two narrow bands each spanning $\sim 10^\circ$. The analytically estimated merger window is overplotted in the solid and dashed lines. 
    In contrast, under octLK, a prograde initial inclination is strongly favored, and both successful migration and disruption can happen over a much wider range. 
    On the other hand, systems originating from small initial orbital periods have similar outcomes under quaLK and octLK, with the latter having a slightly wider formation window.  }
    \label{fig:outcome_vs_inc}
\end{figure}


We now consider the formation outcome as a function of the initial mutual inclination between the two planets in Fig.~\ref{fig:outcome_vs_inc}. 
A clear distinction exists when comparing quaLK (top panel) and octLK (bottom panel). In the quaLK regime, only systems with high initial mutual inclinations $I_{io, {\rm init}}$ $\approx 70-110^\circ$ successfully produced a high-e migration of WASP-107 b. This is because angular momentum is the z-axis $L_z$ is conserved in the test particle quaLK regime~\citep{Naoz:16}, thus a closer-to-perpendicular initial inclination leads to a stronger eccentricity excitation on the inner planet. A formation window forms because the pericenter needs to be small enough to enable migration but large enough to avoid disruption~\citep{Anderson:16, Vick:19}. 
As a reference, we also show the analytically estimated migration window between the solid and dashed lines. They are estimated using the critical values of $r_p$ respectively given in Eq.~(\ref{eq:r_p_mig}) and (\ref{eq:r_p_dis}) with $c_{\rm dis}=2.7$, which are further converted to allowed range on $I_{io,{\rm init}}$ via Eq.~(\ref{eq:c_io_roots}). 
Because of the inner orbit's back-reaction on the outer one, the window is symmetric about $92^\circ$ instead of $90^\circ$ (Eq. \ref{eq:c_io_lim}).  


With octLK, however, significant eccentricity excitation can happen over a wider range of initial mutual inclinations \citep{Liu:15, Petrovich:15}. WASP-107 b can be produced by high-e migration over a wider range of initial mutual inclination $I_{io, {\rm init}}$ $\approx 50-70^\circ$ between planet b and planet c. This more moderate initial mutual inclination can be produced by a less extreme dynamical process such as planet-planet scattering between planet c and/or an undetected third planet \citep{Ford, Lu:24}, or maybe a stellar flyby \citep{Batygin}.

\subsection{WASP-107 b most likely formed within the snowline}
\label{sec:pop_a_init}

\begin{figure}
    \centering
    \includegraphics[width=1\linewidth]{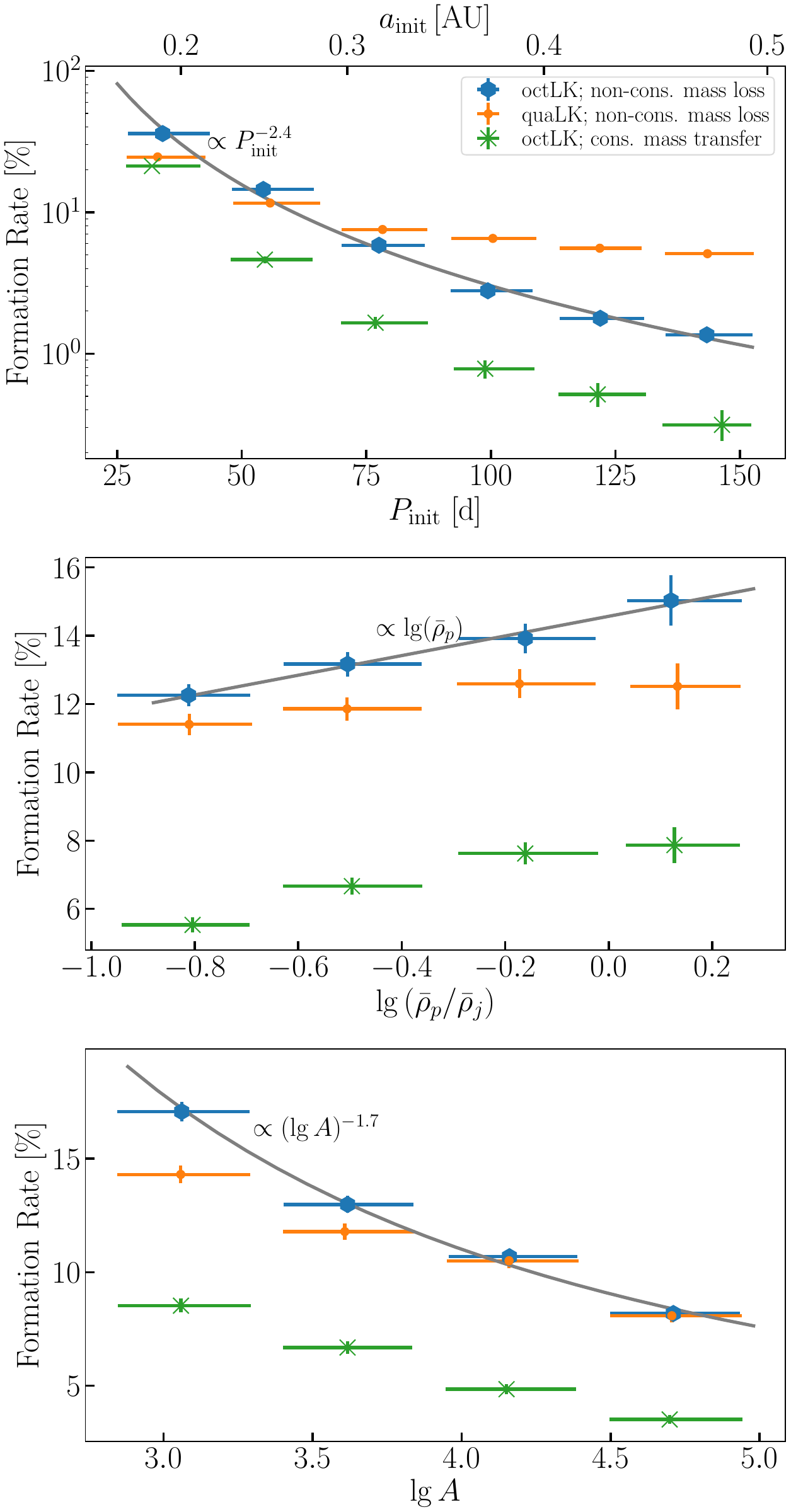}
    \caption{Formation rate of WASP-107b-like systems as functions of the initial inner semi-major axis (top panel; also shown in the top axis is the initial orbital period), initial planetary density (middle panel), and the phenomenological parameter $\lg A$ (bottom panel). Error bars in the y-axis are the 10th and 90th percentiles assuming the number in each bin follows a Poisson distribution with the mean value given by the numerical experiment. 
    Different colors represent different prescriptions of the synthesis (see the legend for details). Phenomenological fits to the case of octLK with nonconservative mass loss are shown in gray lines. Overall, the formation rate favors systems that experience a small amount of mass loss per orbit (small $r_t$ and small $A$). }
    \label{fig:formation_rate}
\end{figure}

As shown in the previous section, tidal disruption is the most likely outcome of high-e migration if the inner planet ventures too close to the host star. WASP-107 b has to have an initial orbital configuration that avoids tidal disruption. This favors a smaller initial semi-major axis for the inner planet (top panel of Fig.~\ref{fig:formation_rate}). This is because at a given $I_{io, {\rm init}}$, a smaller initial semi-major axis provides a small lever arm for the outer planet to act on, thereby limiting the eccentricity excitation through the LK effect (see, Eqs.~\ref{eq:c_io_vs_j_min} and \ref{eq:j_lim_ST}). In other words, the orbital eccentricity of planet b does not reach extreme values that would disrupt the planet during migration. This can also be seen by examining the merger windows in Eq.~(\ref{eq:r_p_mig}) and (\ref{eq:r_p_dis}). Note specifically that $r_{p,{\rm mig}}\propto a_{\rm init}^{-1/7}$ whereas $r_{p,{\rm dis}}$ is independent of $a_{\rm init}$. Consequently, the allowed range of $r_p$ (hence the range of $I_{io,{\rm init}}$) narrows as $a_{\rm init}$ increases, which can also be seen from Fig.~\ref{fig:outcome_vs_inc}. 
We found empirically that the relative formation rate of WASP-107 b in our population synthesis (agnostic about the prior) decreases as $\propto P_{\rm init}^{-2.4}$ (or $\propto a_{\rm init}^{-3.6}$; gray line in Fig.~\ref{fig:formation_rate}). The formation rate is higher for octLK than quaLK if the initial period is less than $\approx70\,{\rm d}$ as the former has a wider formation window (see later in Fig.~\ref{fig:outcome_vs_inc}). For greater initial periods, octLK has a lower formation rate because more significant eccentricity excitation increases the disruption rate. If the mass loss is largely non-conservative (which is more relevant to the numerical simulations of \citealt{Guillochon:11}), the formation rate is 2-3 times higher than runs assuming conservative mass transfer.  

As we will see later in Sec.~\ref{sec:pop_tidal_diss}, formation at large $a_{\rm init}$ is further disfavored if the tidal lag time $t_{\rm lag}$ is smaller than the assumed $t_{\rm lag}=10\,{\rm s}$. This can again be seen from Eq.~(\ref{eq:r_p_mig}) that $r_{r, {\rm mig}}$ decreases with a decreasing $t_{\rm lag}$, making the allowed window narrower.

WASP-107 b most likely did not start with a semi-major axis beyond 0.5 AU. This is not only due to the quickly declining relative formation rate (Fig.~\ref{fig:formation_rate}), but also because a larger initial orbit for planet b would lead to orbit crossing between planet b and planet c in our simulations. During the LK oscillation, the inner orbit will reach an eccentricity $e\simeq 1$, and the inner planet's apocenter will be at $\simeq 2 a_{\rm init}$. Meanwhile, the outer orbit has an eccentricity $0.28$ today. Its value needs to be greatly initially because during the LK oscillation as the outer orbit tends to extract angular momentum from the inner one.  An initial inner semi-major exceeding 0.5 AU will therefore cause the inner apocenter to be comparable or greater than the outer pericenter. 

The location of the snowline for a 0.6-$M_\odot$ K-star is uncertain and most likely time-dependent, but probably near or above 1AU; see fig. 1 of \citet{Kennedy:08}, and also \citep{Ida_Lin}. This suggests that before the onset of high-e migration, WASP-107 b might have formed or migrated within the snowline. \citet{Welbanks:24} reported an atmospheric metallicity that is 10-18 $\times$ solar, and a carbon-to-oxygen ratio C/O = 0.33$^{+0.06}_{-0.05}$. \citet{Sing} reported even higher planetary atmospheric metallicity of 43$\pm$8 times solar. The high metallicity is consistent with a low-mass progenitor for WASP-107 b when considering the mass-metallicity correlation of solar-system planets and exoplanets \citep{Welbanks}. The low C/O ratio is also consistent with forming WASP-107 b within the water snowline \citep{Oberg} as our simulations suggest. More sophisticated models that consider the locking of O into silicates also generally agree that planets which formed within the snowline should have a low C/O ratio \citep[see e.g.][]{Chachan}

\subsection{Final inclination, obliquity, and orbital period}

\begin{figure}
    \centering
    \includegraphics[width=1\linewidth]{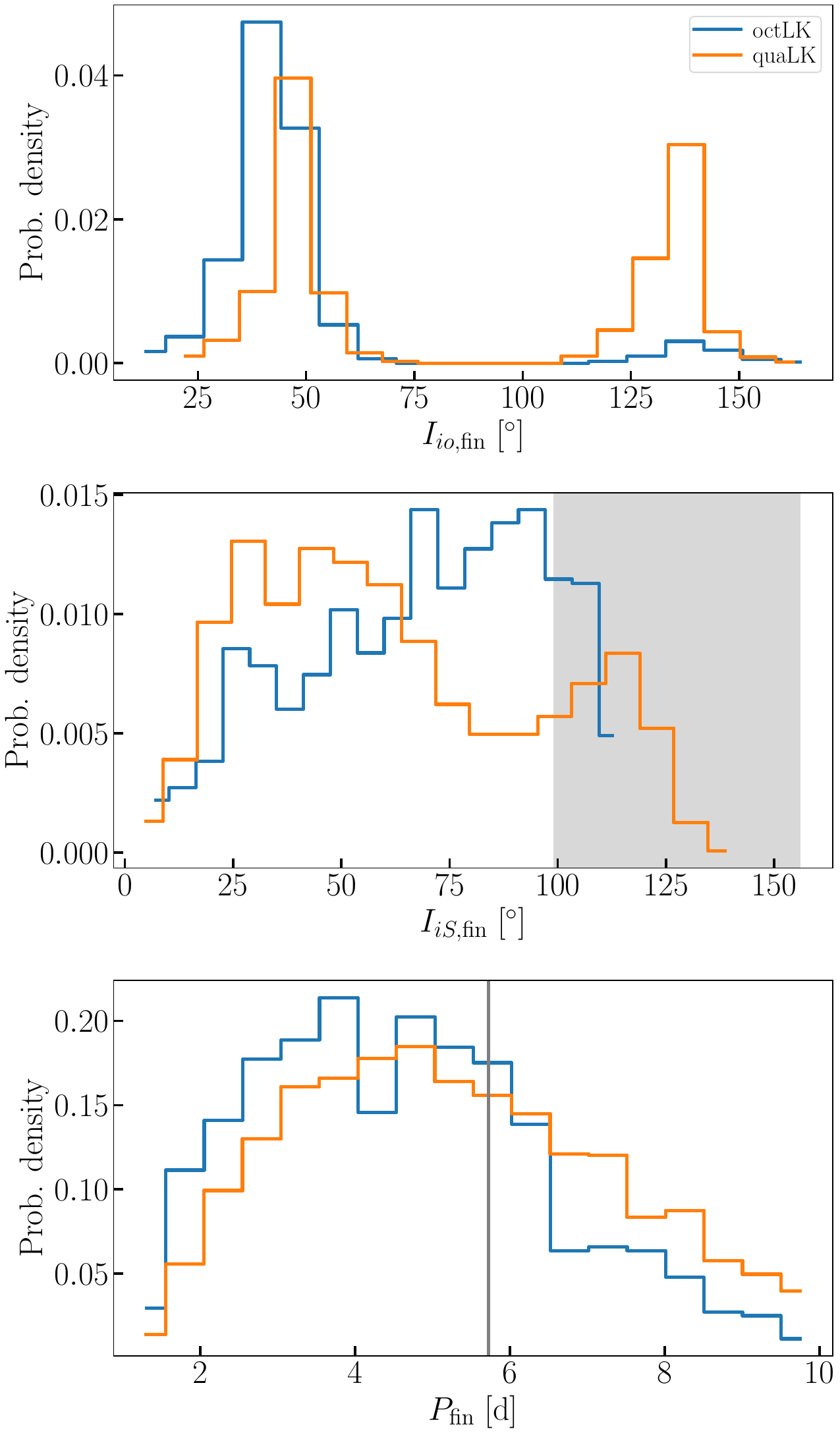}
    \caption{Distributions of the final properties for systems with initial period $P_{\rm init}>75\,{\rm d}$ and non-conservative mass loss. Top: mutual inclination between inner and outer orbits. 
    A moderate mutual inclination around $25-50^\circ$ is favored under octLK. 
    Middle: distribution of the final stellar obliquity. Both cases support a wide distribution of obliquities and are consistent with the observed near polar configuration of WASP-107 b. This is consistent with the high inferred obliquity of WASP 107b. The measured value (including uncertainties) is shown in the gray band.
    Bottom: final period. A 3-6 d final period is more favored. The gray vertical line shows the observed period (with uncertainty smaller than linewidth).
    }
    \label{fig:fin_dist}
\end{figure}

\begin{figure}
    \centering
    \includegraphics[width=1\linewidth]{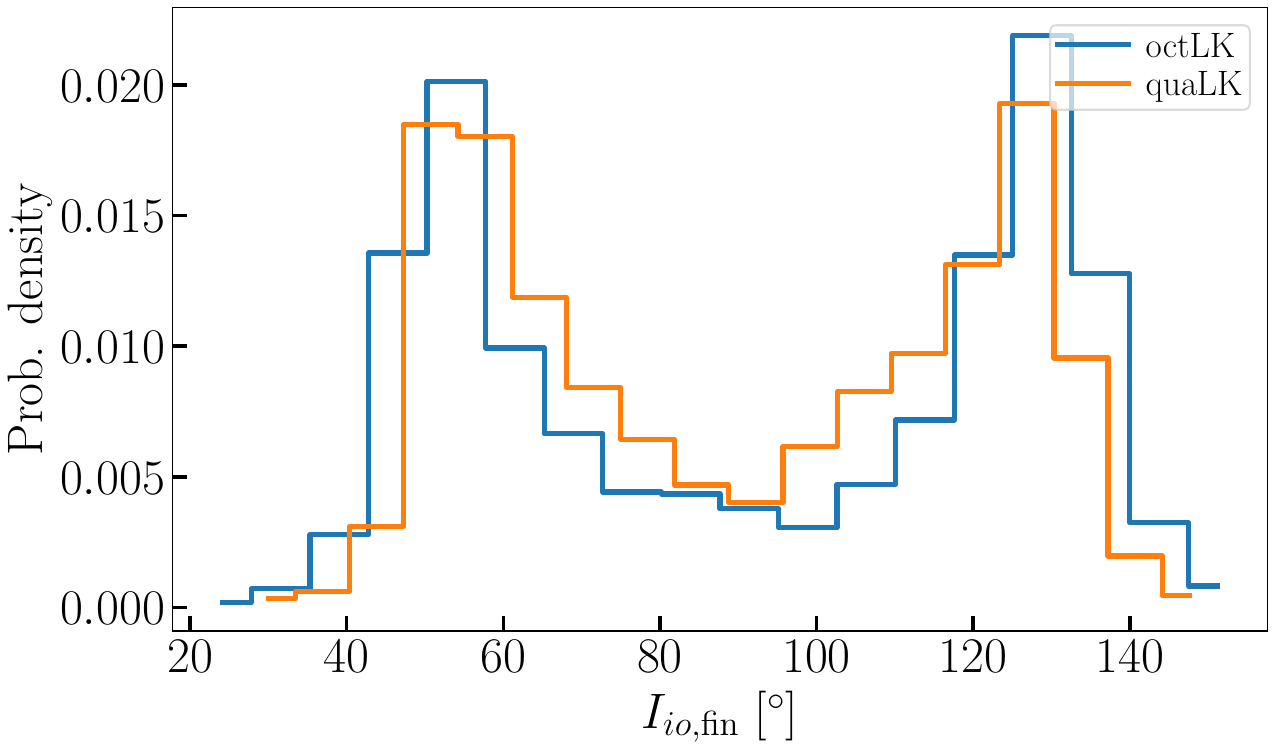}
    \caption{Similar to Fig.~\ref{fig:fin_dist} but for systems with smaller initial period, $25\,{\rm d}\leq P_{\rm init}<40\,{\rm d}$. The octLK and quaLK simulations now show similar distributions spanning $40-140^\circ$. The obliquity and final period distributions are similar to the quaLK results in Fig.~\ref{fig:fin_dist} and are not repeated here. 
    }
    \label{fig:fin_dist_2}
\end{figure}

We now compare the properties of the successfully migrated systems with the observed properties of WASP-107. In Figs.~\ref{fig:fin_dist} and \ref{fig:fin_dist_2}, we show the final mutual inclination between the inner and outer planets $I_{\rm io, fin}$  (top panel), the stellar obliquity $I_{iS}$ with $\cos I_{iS}=\uvect{J} \cdot \uvect{S}_\ast$ (middle panel), and the final orbital period of planet b $P_{\rm fin}$. Fig.~\ref{fig:fin_dist} is for systems originating from a wide initial orbits, $P_{\rm init} > 75\,{\rm d}$, and Fig.~\ref{fig:fin_dist_2} for those with $25\,{\rm d}\leq P_{\rm init}<40\,{\rm d}$. 

Systems evolved under quaLK show roughly even distribution with a final mutual inclination peaking at either $I_{\rm io, fin} = 45^\circ$ or $135^\circ$, insensitive to the initial orbital period. However, the distributions change significantly for octLK systems when the initial period varies. Whereas those started with small $P_{\rm init}<40\,{\rm d}$ show a distribution similar to the quaLK cases (Fig.~\ref{fig:fin_dist_2}), systems originated from $P_{\rm init}>75\,{\rm d}$ (top panel of Fig.~\ref{fig:fin_dist}) favor a lower mutual inclination $I_{io, {\rm fin}} \approx 25-50^\circ$. If measured with {\it Gaia} astrometry (\citealt{Gaia};  a similar measurement has been done on HAT-P-11 \citealt{Xuan}), the mutual inclination can hence be used to constrain $P_{\rm init}$. An $I_{io,{\rm fin}}< 40^\circ$ would suggest $P_{\rm init}>75\,{\rm d}$, whereas an $I_{io,{\rm fin}}> 50^\circ$ would point to an initial period $P_{\rm init}<40\,{\rm d}$.

The stellar obliquity $I_{iS}$ is shown in the middle panel of Fig.~\ref{fig:fin_dist}. In both cases, the final stellar obliquity encompasses $110^\circ$ which is the empirically measured stellar obliquity \citep{Rubenzahl}. LK consistently launches the inner planet into misaligned orbit around the host star thanks to the dynamical attractor effect described in \citet{Liu:18}. Similar obliquity distribution is also observed in \citet{Angelo:22} for Kepler-1656b (see, e.g., their fig. 11 for the tidally locked systems). In all cases, the orbital period distributions peak between 3-6 d, also consistent with the period of WASP-107 b.

\subsection{Migration Timescale}

\begin{figure}
    \centering
    \includegraphics[width=1\linewidth]{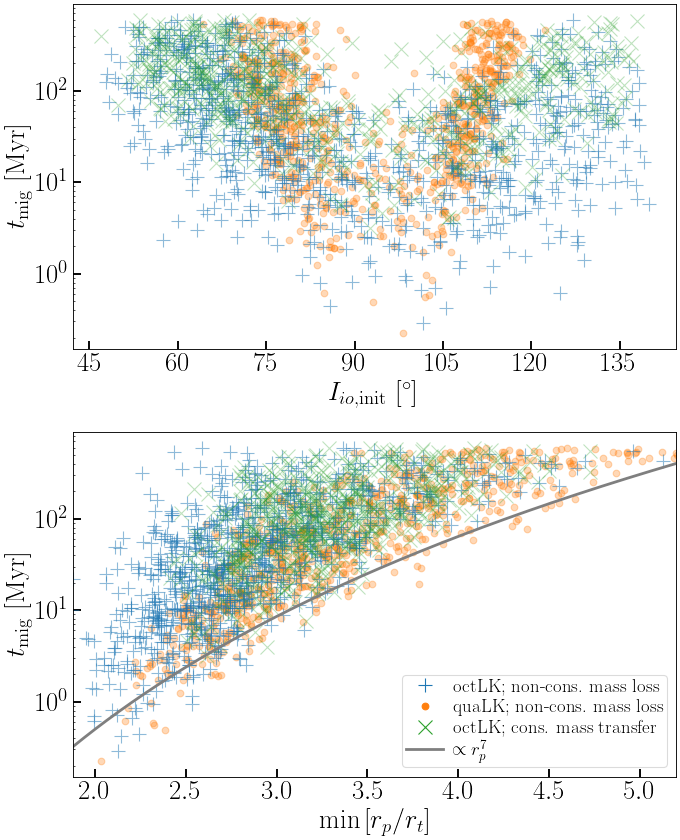}
    \caption{Migration time versus initial inclination between inner and outer orbits (top) and the minimum pericenter separation during the LK cycle (bottom). Different colors and marker styles indicate different orders in the LK evolution and different prescriptions of mass loss/transfer (see legend). 
    }
    \label{fig:t_mig}
\end{figure}

In Fig. \ref{fig:t_mig}, the migration time, measured from the onset of LK oscillation to the point when the eccentricity drops below 0.1 and period below 10 d, is shown as a strong function of the pericenter separation $t_{\rm mig}\propto r_p^7$. This is because $a/\dot{a}_t\propto r_p^{15/2}$ in the adopted constant time lag tidal models~\citep{Hut:81}, and a planet spends a fraction of $\sim \sqrt{1-e^2}$ of time during the LK cycle at the highly eccentric stage \citep{Anderson:16}. A secondary effect is that at a given value of ${\rm min}[r_p/r_t]$, systems evolved under octLK require longer times to migrate than those under quaLK. This is because the eccentricity oscillation under octLK is more stochastic and occasionally reach down to a very small pericenter while the mean separation is greater. 

As explained in Sec.~\ref{sec:mutual_inclination}, the initial mutual inclination between planet b and c often determines the minimal pericenter separation ${\rm min}[r_p/r_t]$. Therefore, the initial mutual inclination also changes the migration timescale (upper panel of Fig. \ref{fig:t_mig}). Systems with more moderate initial mutual inclinations generate gentler, slower high-e migrations.

\subsection{Impact of the tidal dissipation efficiency}
\label{sec:pop_tidal_diss}

\begin{figure}
    \centering
    \includegraphics[width=1\linewidth]{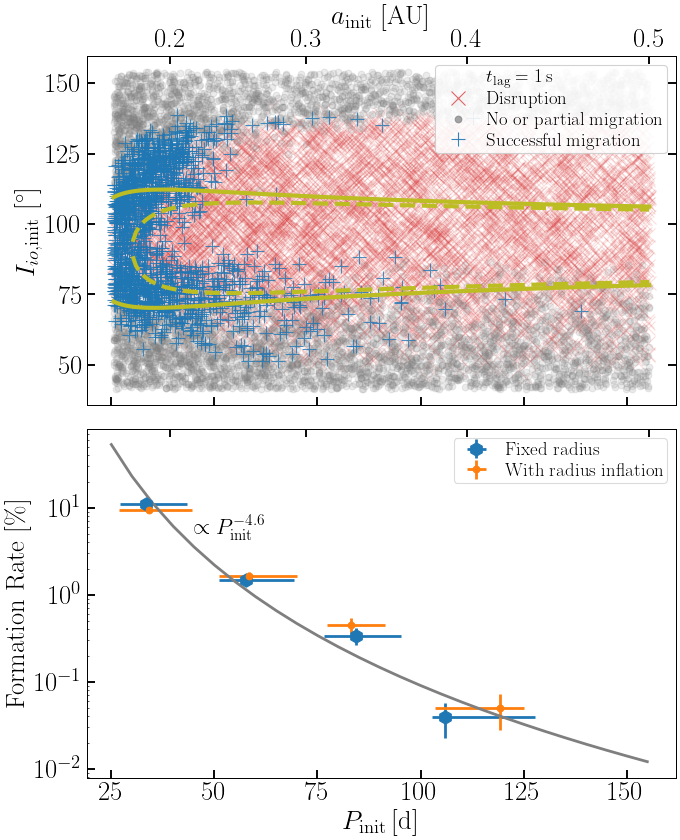}
    \caption{Formation window and rate of WASP-107-like systems with $t_{\rm lag}=1\,{\rm s}$. The top panel is similar to the lower panel of Fig.~\ref{fig:outcome_vs_inc}. The markers are computed for octLK systems, and the solid and dashed lines represent analytically estimated merger window under the quadrupole approximation. 
    The bottom panel is similar to the top panel of Fig. \ref{fig:formation_rate}. 
    Including radius inflation or not does not change the formation window and rate significantly.
    }
    \label{fig:tau_1s_results}
\end{figure}

While we adopted $t_{\rm lag}=10\,{\rm s}$ as the default value for the tidal lag time, we also consider the case where $t_{\rm lag}=1\,{\rm s}$ for completeness. The choice of $t_{\rm lag}=1\,{\rm s}$ was also adopted by previous analyses including \citet{Anderson:16,Vick:19}. For this section, we focus on octLK with non-conservative mass loss from the planet.

The main impact of reducing $t_{\rm lag}$ from 10\,s to 1\,s, as illustrated in Fig.~\ref{fig:tau_1s_results}, is that the overall formation rate is reduced by about a factor of 4. Furthermore, very few systems with $P_{\rm init}\gtrsim 100\,{\rm d}$ can form with reduced $t_{\rm lag}$ as shown in the top panel. The formation rate also decays faster with respect to $P_{\rm init}$ as the gray line in the bottom panel depicts (cf. the top panel of Fig.~\ref{fig:formation_rate}). This behavior can be understood by again examining the formation window determined by $r_{p,{\rm dis}} < r_p < r_{p, {\rm mig}}$. The disruption threshold (Eq.~\ref{eq:r_p_dis}) is independent of both $t_{\rm lag}$ and $a_{\rm init}$, whereas the migration threshold decreases with both decreasing $t_{\rm lag}$ and increasing $a_{\rm init}$. Indeed, the analytically estimated merger window (under the quadrupole approximation) between the solid and dashed line in the top panel essentially vanishes for $P_{\rm init} \gtrsim 100\,{\rm d}$. Therefore, a smaller dissipation would strengthen the point we made in Sec.~\ref{sec:pop_a_init} that  WASP-107 b likely started the LK oscillation within the snowline.

With $t_{\rm lag}=1\,{\rm s}$, there are still systems in our simulation that experience 10-20\% radius inflation and have $L_t>10^{-7} L_\odot$ near the end of the evolution, yet they typically have $P_{\rm fin}\lesssim 3\,{\rm d}$ to reach the desired tidal heating rate. 
Similar to the case where $t_{\rm lag}=10\,{\rm s}$, including the radius inflation or not affects mainly the final planetary radius but not other properties of the system, as demonstrated in the bottom panel of Fig. \ref{fig:tau_1s_results}. In both cases, the final obliquity ($I_{iS, {\rm fin}}$) and orbital inclination ($I_{io, {\rm fin}}$) are similar to the quaLK case shown in Fig.~\ref{fig:fin_dist}, both supporting polar or even retrograde orbits. 

\section{Conclusions}
\label{sec:conclusions}

We conclude by answering the questions we posed in the Introduction:

\begin{itemize}
     \item The observed WASP-107 c is indeed capable of initiating a high-e migration for WASP-107 b. The initial mutual inclination between the two planets has to be nearly orthogonal (70-110$^\circ$) in the case of quadrupole LK evolution. If including the octupole LK terms, the eccentricity excitation is more stochastic, and the initial mutual inclination can be as low as 50$^\circ$ which may be more easily produced by an earlier dynamical process. Our simulations are agnostic about how the initial mutual inclination emerged. It could be a planet-planet scattering \citep{Chatterjee:08,Ford}, or maybe a stellar flyby \citep{Batygin}. The reader is also referred to the disk-driven resonance scenario suggested by \citep{Petrovich:20}.
     We further predict the current-day mutual inclination between the two planets to be $25-50^\circ$ should the planet originate from an initially wide orbit with a period greater than 75 days, or $40-140^\circ$ should the initial period be less than 40 days. Future {\it Gaia} observations may help to distinguish the two scenarios.

    \item High-e migration can indeed explain the polar orbit of WASP-107 b \citep{Dai_2017,Rubenzahl}: the final stellar obliquity in our simulations encompasses 110-120$^\circ$ that is reported in the literature.
    
    \item Our simulations suggest that WASP-107 b is still circularizing and tidally decaying on its current-day 5.7-day , $e = 0.06$ orbit. With a Neptune-like tidal quality factor, the tidal heating can be as high as $10^{-6} L_\odot$, which is consistent with the observed high intrinsic temperature \citep{Welbanks:24,Sing}.

    \item Our simulations suggest that WASP-107 b most likely started high-e migration within 0.5 AU. Orbital crossing and tidal disruption are the more likely outcomes if the planet has a wider initial orbit. In other words, WASP-107 b most likely formed or migrated within the snowline before high-e migration. The observed low C/O = 0.33 \citep{Welbanks:24} seems consistent with a formation within the water snowline.

    \item Our simulations suggest that WASP-107 b most likely started with a mass no more than 20\% greater than its current mass. In other words, WASP-107 b did not start as a Jupiter-mass planet. Stronger mass loss during high-e migration usually initiates a run-away tidal disruption. The fractional mass loss also strongly correlates with the migration time. If the system took 600 Myr (estimated age) to migrate, the mass loss would be minimal.  20\% fractional mass loss corresponds to a fast migration of only tens of Myr. 

    \item WASP-107 b's radius may be only inflated during the final stage of high-e migration where both strong stellar insolation and tidal heating contributed to the inflation of the observed Jupiter-like radius. 

\end{itemize}

Putting pieces together, we propose the following potential formation scenario of WASP-107 b (and planets with similar orbital architectures). The progenitor planet has an initial mass between 0.1-0.2 $M_j$ (Fig.~\ref{fig:dMp}). Before the onset of LK, it might have formed or migrated within the water snowline with a semi-major axis of $<$0.5\,{\rm AU} (top panel of Fig.~\ref{fig:formation_rate}). A Jupiter-like outer planet (WASP-107 c) on a moderately inclined (top panel of Fig.~\ref{fig:t_mig}), eccentric orbit excites the inner planet to a highly eccentric orbit via octupole LK oscillations. WASP-107 b should have a smaller radius of around $0.7 R_j$ before high-e migration which helps it survive tidal disruption (middle panel of Fig.~\ref{fig:formation_rate}). Yet near the end of the LK cycle, strong tide heating may coupled with strong stellar insolation, inflating its radius to around $R_j$ (second-last panels of Figs.~\ref{fig:wasp_107b_sample_traj_2_2731} and \ref{fig:wasp_107b_sample_traj_9_994}). After this inflation phase, the radius may stay inflated if the migration finished recently (less than the deflation timescale).
Today, the planet is still on a misaligned, eccentric orbit. The resultant eccentricity tides cause internal heating, give a high intrinsic temperature, and partially drives the observed hydrodynamic mass loss of the planet.

During the preparation of this manuscript, \citet{Lu:24} conducted a similar analysis on the HAT-P-11 system. They performed N-body simulations which can explain the initial production of a significant mutual inclination between planets b and c necessary to trigger the LK oscillation. An N-body simulation can also better address the stability of orbital crossing systems which we discarded in this study. On the other hand, our study allows for more flexibility in the planet's evolution (including mass loss and delayed radius evolution). A future work synthesizing the study by \citet{Lu:24} and ours will be worth investigating. 
Besides the mechanism discussed in this work, other mechanisms (e.g., disk-driven resonance; \citealt{Petrovich:20}) may also be able to explain some observed properties of WASP-107 b and similar systems. More detailed comparisons between different mechanisms at a population level should be conducted in future studies. 

\section*{Software availability}
Source code to generate results presented in this work are publicly available on GitHub at \url{https://github.com/hangyu45/HighEccentricityMigration} under a GPL-3.0 License.

\section*{Acknowledgments}
We thank Heather Knutson, Yayaati Chachan, Tiger Lu, Gudmundur Stefansson, and Ryan Rubenzahl for their helpful discussions. We thank Luis Welbanks, Gongjie Li, and Tiger Lu for sharing their manuscripts with us early.
Computational efforts were performed on the Tempest High Performance Computing System, operated and supported by University Information Technology Research Cyberinfrastructure at Montana State University. H.Y. acknowledges support from NSF grant No. PHY-2308415.


\bibliography{ref}{}
\bibliographystyle{aasjournal}



\end{document}